\documentclass[10pt,twocolumn,twoside]{IEEEtran} % Use this for two-column version
\usepackage{graphicx,amssymb,epstopdf,amsmath,cite,subcaption,array, tikz,verbatim,amsthm,dsfont,multirow,tablefootnote}
\usetikzlibrary{arrows,chains,matrix,positioning,scopes}

\newtheorem{thm}{Theorem}[section] %(If you want theorem numbered
\newtheorem{lemma}{Lemma}[section] %%    with section number.
\newtheorem{cor}{Corollary}[section]
\newtheorem{prop}{Proposition}[section]
\newtheorem{definition}{Definition}[section]

\newcommand{\afh}[3]{\mathcal{H}(#1,#2,#3)}
\newcommand{\mc}[4]{\mathcal{C}(#1,#2,#3,#4)}
\newcommand{\dtft}[2]{#1(e^{j2\pi f #2})}
\newcommand{\boldvec}[1]{\boldsymbol{#1}}

\title{Reconstruction of Frequency Hopping Signals \\ From Multi-Coset Samples}
\author{Chia Wei Lim and Michael B. Wakin%
\thanks{CWL is with DSO National Laboratories of Singapore; e-mail: lchiawei@dso.org.sg. MBW is with the Department of Electrical Engineering and Computer Science, Colorado School of Mines; e-mail: mwakin@mines.edu. This work was partially supported by DSO National Laboratories of Singapore and by NSF grant CCF-1409258.}
\date{}
}

\begin{document}

\maketitle

\begin{abstract}
Multi-Coset (MC) sampling is a well established, practically feasible scheme for sampling multiband analog signals below the Nyquist rate. MC sampling has gained renewed interest in the Compressive Sensing (CS) community, due partly to the fact that in the frequency domain, MC sampling bears a strong resemblance to other sub-Nyquist CS acquisition protocols. In this paper, we consider MC sampling of analog frequency hopping signals, which can be viewed as multiband signals with changing band positions. This nonstationarity motivates our consideration of a segment-based reconstruction framework, in which the sample stream is broken into short segments for reconstruction. In contrast, previous works focusing on the reconstruction of multiband signals have used a segment-less reconstruction framework such as the modified MUSIC algorithm. We outline the challenges associated with segment-based recovery of frequency hopping signals from MC samples, and we explain how these challenges can be addressed using conventional CS recovery techniques. We also demonstrate the utility of the Discrete Prolate Spheroidal Sequences (DPSS's) as an efficient dictionary for reducing the computational complexity of segment-based reconstruction.
\end{abstract}

\section{Introduction}
\label{sec:mc_dpss_introduction}

Frequency hopping (FH) signals, which arise in spread-spectrum multiple access (SSMA) communication systems, have long been used in military radios and more recently appear in the Bluetooth protocol as a form of communication offering excellent anti-jam capability. An FH signal rapidly changes its transmission frequency over a bandwidth that is much larger than its original bandwidth. Over time, the FH signal can be viewed as a concatenation of multiple short duration bursts (hops) each having a frequency that is pseudo-randomly chosen from a preselected set of possible hopping frequencies~\cite{handbook_ss,digital_comms_proakis}.

Typically, accurate synchronization is required amongst cooperative FH receivers (having a priori knowledge of hopping frequencies) in a network. However, in the  absence of such knowledge (subsequently referred to as the \emph{blind} setting), constant monitoring over the entire transmission spectrum requires costly wideband receivers with large instantaneous bandwidths. This imposes a severe burden on the analog-to-digital converters (ADCs) used in state-of-the-art digital receivers. With conventional uniform sampling and reconstruction, the Nyquist sampling theorem \cite{nyquist_1} dictates a minimum sampling rate of twice the transmission bandwidth, which can be on the order of tens of gigahertz.

In spite of the high Nyquist limit, the fact that each transmitted hop is narrowband suggests the possibility of sub-Nyquist rate sampling even in the blind setting. As we discuss, several previous works in the literature have considered sub-Nyquist rate sampling of blind \emph{multiband} signals---in particular via Multi-Coset (MC) sampling---but few have considered the blind FH signal model. A multiband signal is one whose spectrum is concentrated on a small number of narrow bands, and the prefix ``blind'' indicates a lack of knowledge of the center frequencies and widths of the signal bands. The fact that the transmission frequency of the blind FH signal changes over time limits the feasibility of previously proposed blind multiband signal recovery techniques, most of which do not require deliberate segmenting of MC sample streams into finite-length segments for signal recovery. In the sequel, we refer to such a recovery framework as {\em segment-less} and discuss a variant of the classical MUSIC algorithm \cite{schmidt_1} for this framework in Section~\ref{sec:mc_dpss_blind_multiband_signal_reconstruction_from_mc_samples}. Due to the non-stationary nature of blind FH signals (changing transmission frequencies), there is a need to consider an intentional segmenting of MC sample streams into segments for signal recovery. In the sequel, we refer to such a recovery framework as {\em segment-based}.

In this paper, we assume there exists a front-end MC sampler, which is explained in Section~\ref{sec:mc_dpss_mc_sampling}, producing sub-Nyquist rate periodic nonuniform samples of a sum of multiple analog FH signals. We highlight a salient aspect of the MC sampler which facilitates segment-based recovery, and we advocate the need for such a recovery framework for the blind analog FH signal model. As we discuss, segment-based recovery, which essentially is a finite dimensional framework, is an appropriate setup in which Compressive Sensing (CS) signal recovery approaches can be used to recover blind analog FH signals from MC samples. Specifically, we show that the resulting linear system can be recast as the classical CS multiple measurement vector (MMV) problem and that CS MMV solvers can used for signal recovery. Further, we propose the use of a Discrete Prolate Spheroidal Sequence (DPSS) vector-based dictionary which reduces the scale of the signal recovery problem, thereby improving solver latency.

\subsection{Blind Analog Frequency Hopping (FH) Signal Model}
\label{sec:mc_dpss_blind_fh_signal_model}

In this section, we introduce the blind analog frequency hopping (FH) signal model. We denote this class of signals as $\afh{N}{B}{T}$, where $N$ is the number of FH radios, $B$ is the maximum bandwidth of each narrowband frequency hop in Hz, and $T$ is the minimum hop repetition interval (HRI) in seconds. Formally, signals $x(t)$ belonging to $\afh{N}{B}{T}$ can be written as
\begin{equation}
\label{eqn:bafh_signal_model_1}
x(t)=\sum^N_{i=1}\sum^{n_i-1}_{k=0}g_{ik}(t-kT_i-\tau_i)e^{j(2\pi f_{ik} t+\theta_{ik})},
\end{equation}
where
\begin{equation*}
%\label{eqn:bafh_signal_model_2}
g_{ik}(t)=r_{i}(t)m_{ik}(t),
\end{equation*}
$n_i$ is the number of hops for the $i$th radio, $r_{i}(t)$ is a time-limited and essentially bandlimited window characteristic of the $i$th radio, $m_{ik}(t)$ is the baseband modulated signal containing the information symbols transmitted by the $i$th radio's $k$th hop, $T_i$ is the $i$th radio's HRI, $\tau_i$ is the delay offset of the $i$th radio used to model the asynchronous transmission nature\footnote{In the synchronous transmission case, all radios have the same $T_i$ and they transmit at the same time such that $\tau_i=\tau\;\forall i$.} of the radios and satisfies $\tau_i\le T_i$, $f_{ik}$ is the carrier frequency of the $i$th radio's $k$th hop, and $\theta_{ik}$ is the carrier phase of the $i$th radio's $k$th hop.

As is evident from~\eqref{eqn:bafh_signal_model_1}, signals belonging to $\afh{N}{B}{T}$ are parameterized by numerous variables characteristic of each FH radio. In particular, our reference to $\afh{N}{B}{T}$ as a ``blind'' FH signal model reflects the fact that we assume prior knowledge of $N$, $B$, and $T$ \emph{only}, where
\begin{equation}
B\ge\sup\{f:|G_{ik}(f)|> \epsilon\},
\label{def:afh_B}
\end{equation}
$G_{ik}(f)$ denotes the continuous time Fourier transform (CTFT) of $g_{ik}(t)$, $\epsilon$ denotes a small positive constant (since $g_{ik}(t)$ is an essentially bandlimited window), and
\begin{equation*}
T\le\min_i T_i.
%\label{eqn:minimum_hri}
\end{equation*}
The hopping frequencies are assumed to satisfy $f_{ik}\in [f_{\text{min}},f_{\text{max}}]$ which corresponds to a continuum of possible hopping frequencies.\footnote{We assume that $f_{\text{min}}$ and $f_{\text{max}}$ are known, as with any other typical signal acquisition setup.} This condition allows the signal model to include hopping frequencies that are commonly referred to as ``off-grid'' and in general, need \emph{not} be an integer multiple of some fundamental frequency $f_0$. The ``on-grid'' assumption---along with an assumed knowledge of $f_0$---are commonly adopted for the sake of simplifying blind FH signal models in the literature.

\begin{figure}[t]
	\centering
	\includegraphics[width=0.5\textwidth]{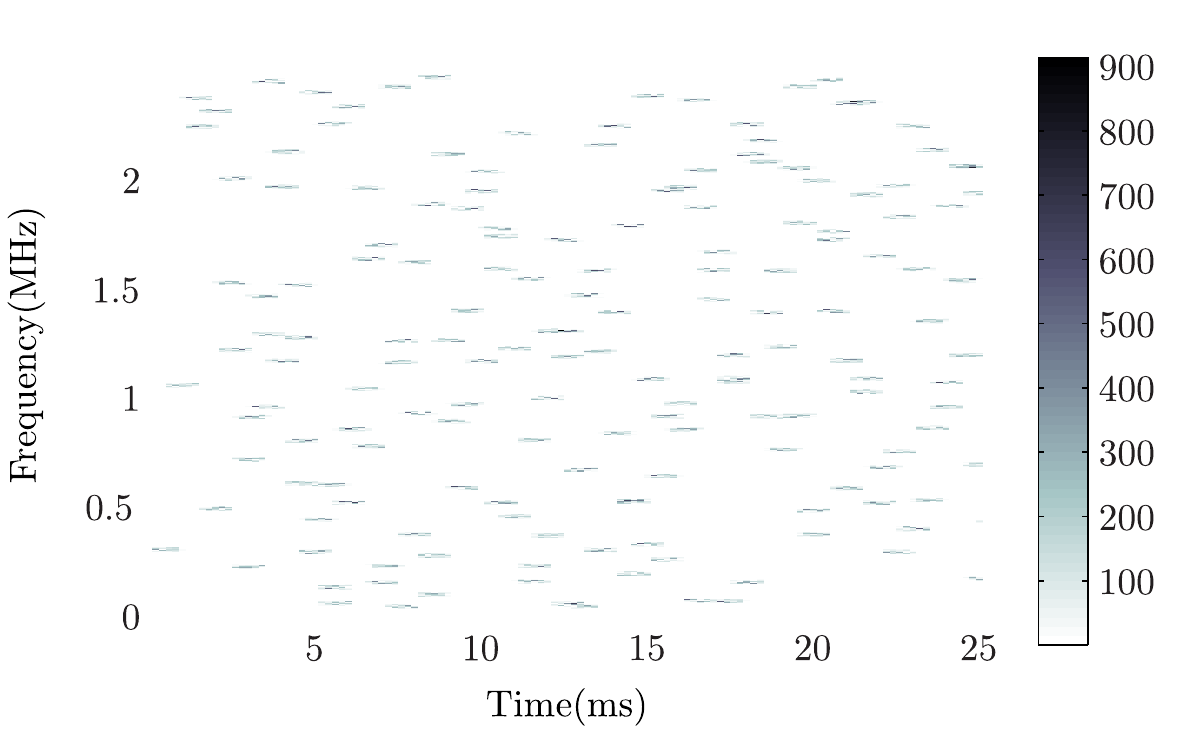}
	\caption{\label{fig:spectrogram_sfh}Spectrogram of a signal belonging to $\afh{7}{25000}{1\times 10^{-3}}$. In this plot, all the emitters have an identical HRI of 1 ms corresponding to a hop rate of 1000 hops/s.}
\end{figure}

A plot of the spectrogram of a signal belonging to $\afh{7}{25000}{1\times 10^{-3}}$ is shown in Fig.~\ref{fig:spectrogram_sfh}.

\subsection{Previous Works}
\label{sec:mc_dpss_previous_works}

Among the ample literature available with regards to sub-Nyquist rate MC sampling and reconstruction of multiband signals, only literature relevant to this paper is mentioned in the sequel. The interested reader is referred to \cite{landau,1053794,103094,544131,560608,pfeng_phd,868487,950786,4601017,4749297,5743619,davenport2012compressive,6621277} for more information on related works.

While the Nyquist sampling theorem holds for any bandlimited signal, Landau \cite{landau} showed that the required average (both uniform and nonuniform) sampling rate for multiband signals can be reduced to the Landau rate, which is the Lebesgue measure of the signal's spectral support. Early works on nonuniform sampling and reconstruction of bandlimited signals \cite{1053794,103094} and multiband signals \cite{560608} required a priori knowledge of the signal's underlying spectral support.

Feng and Bresler \cite{544131,pfeng_phd} addressed the nonuniform sampling and reconstruction of multiband signals in the blind setting using sub-Nyquist rate MC samples. Reconstruction was achieved by first using a modified MUSIC algorithm to obtain the signal's unknown spectral support and subsequently using standard least squares to recover the multiband signal. It was shown that such a sampling scheme can approach the Landau rate asymptotically.

Venkataramani and Bresler \cite{868487} derived explicit multiband signal reconstruction formulas using sub-Nyquist rate MC samples and derived error bounds on the peak error and the energy of the aliasing error (due to multiband signal modeling mismatch) for the sampling/reconstruction system. Further analysis of the system performance in the presence of additive noise was also provided. Using the previously computed bounds as performance measures, it was subsequently shown \cite{950786} that optimizations such as optimal sampling pattern design and optimal base sampling rate can further improve reconstruction performance.

Bresler \cite{4601017} discussed the connections between the previous works on blind spectrum sensing and those of CS in particular for the multiband signal model. It was argued that blind spectrum sensing using the MC sampler provides efficient sub-Nyquist rate sampling with reconstruction costs linear in the amount of data and with robustness to noise.

Mishali and Eldar \cite{4749297} reexamined the sampling of multiband signals using sub-Nyquist MC samples. It was shown that no sampling scheme can have a worst case performance better than the MC sampler for the multiband signal model. In contrast to previous blind spectrum sensing works, the data covariance matrix was first used to obtain the subspace containing the multiband signal and subsequently a CS-type solver was used to recover the support.

Kochman and Wornell \cite{5743619} investigated the impact of a finite sensing interval on the recovery of the multiband signals' underlying spectral sparsity structure (from sub-Nyquist MC samples) using finite sharp transition band windows and characterized the associated redundancy.

Davenport and Wakin \cite{davenport2012compressive} addressed the sub-Nyquist rate sampling and reconstruction of analog multiband signals using a modulated DPSS based dictionary for efficiently representing the sampled multiband signals. Instead of the MC sampling protocol, that work considered taking random linear projections of Nyquist rate samples of the multiband signal.

Liu et al. \cite{lfeng_1} proposed a CS framework for the interception of FH signals using conventional CS random projections without knowledge of the hopping sequence of the underlying FH signal. In contrast to classical CS applications where one seeks to reconstruct the FH signal, the goal of \cite{lfeng_1} was to perform classification in the compressive domain. Subsequently, in \cite{lfeng_2}, a compressive detection strategy for FH signals was proposed with theoretical detection and false alarm rates derived from previous scanning spectrum analyzer results. Then, in \cite{lfeng_3},  the previously proposed compressive detection strategy was extended to multiple FH signals. We note that the FH ``on-grid'' signal model was used in all three papers.

A comparison of all pertinent works against this paper can be found in Table~\ref{tab:compare_works}. In this table, RP denotes Random Projections, MB denotes the blind multiband signal model, FH-On denotes the FH ``on-grid'' signal model, FH-Off denotes the FH ``off-grid'' signal model, Cov denotes the use of the data covariance matrix for support recovery, Cov+CS denotes the use of the data covariance matrix and a CS-type solver for support recovery, and CS denotes the use of a CS-type solver for support recovery. We note that no preferred support recovery technique was indicated or used in \cite{5743619} although the three recovery techniques were mentioned as possible ways to recover the signal support.

\begin{table}[ht]
\begin{center}
	\caption{\label{tab:compare_works}Comparison of pertinent previous works against this paper in terms of sampling scheme, signal model, and support recovery approach.}
	\scalebox{0.8}{\begin{tabular}{c|c|c|c|c|c|c|c|c|}
		\cline{2-9}
		%&
		& \multicolumn{2}{ c| }{Sampling Scheme} & \multicolumn{3}{ c| }{Signal Model}& \multicolumn{3}{ c| }{Support Recovery}\\ \cline{2-9}
		%&
		& MC & RP & MB & FH-On&FH-Off& Cov& Cov+CS & CS\\ \cline{1-9}
		%\multicolumn{1}{ |c }{\multirow{5}{*}{{Previous works} }} &
		\multicolumn{1}{ |c| }{\cite{544131,pfeng_phd,868487,950786}} & $\times$ & & $\times$ & &&$\times$&&     \\ \cline{1-9}
		%\multicolumn{1}{ |c  }{}                        &
		\multicolumn{1}{ |c| }{\cite{4749297}} & $\times$ &  & $\times$ & &&&$\times$&     \\ \cline{1-9}
		%\multicolumn{1}{ |c  }{}                        &
		\multicolumn{1}{ |c| }{\cite{5743619}} & $\times$ &&$\times$&&&\multicolumn{3}{ c| }{$\times$}     \\ \cline{1-9}
		%\multicolumn{1}{ |c  }{}                        &
		\multicolumn{1}{ |c| }{\cite{davenport2012compressive}} &&$\times$&$\times$& &&&&$\times$     \\ \cline{1-9}
		%\multicolumn{1}{ |c  }{}                        &
		\multicolumn{1}{ |c| }{\cite{lfeng_1,lfeng_2,lfeng_3}} &&$\times$&&$\times$ &&&&$\times$     \\ \cline{1-9}
		%\multicolumn{1}{ |c }{{This paper}}                        &
		\multicolumn{1}{ |c| }{This paper} &$\times$&&& &$\times$&&&$\times$     \\ \cline{1-9}
	\end{tabular}}
\end{center}
\end{table}

\subsection{Need for a Segment-Based Reconstruction Framework}
\label{sec:mc_dpss_need_for_fd_reconstruction_framework}

In a typical idealized analysis of an MC sampling system, signal acquisition and reconstruction take place over an infinite-length sub-Nyquist sample stream. In practical finite systems, of course, such an infinite stream of samples cannot be acquired or processed. Over suitably long segments of samples in time, one might hope to achieve signal reconstruction performance that is comparable to what could be achieved using an infinite segment size; as mentioned previously, \cite{5743619} studies the implications of a finite segment size on multiband signal reconstruction from MC samples.\footnote{We note that the differences between our work and that of \cite{5743619} include the segment-based recovery framework, the $\afh{N}{B}{T}$ signal model, and the application.} In some applications, one could simply increase the segment length (capturing samples over a longer time duration) to achieve the desired quality of reconstruction. In particular, when considering a multiband signal, the spectral support of the signal (and thus the signal's Landau rate) remains fixed over any time duration.

Unfortunately, for FH signals observed over multiple hops, the spectral support grows approximately linearly with the time duration over which the signal is observed. Therefore, successful FH signal reconstruction at a minimum average sampling rate requires a processing segment size during which one expects the spectral support to have minimal growth. In particular, it is necessary to intentionally partition the sample stream into segments that have duration commensurate with the minimum HRI $T$. To deal with these short-duration sample streams, we develop a signal reconstruction framework that is explicitly segment-based.

\subsection{Paper Organization}
\label{sec:paper_organization}

In Section~\ref{sec:mc_dpss_preliminaries} we first review the MC sampling protocol and the segment-less, blind recovery of multiband signals from sub-Nyquist MC samples. We then outline various CS signal recovery concepts and discuss useful properties of DPSS vectors.

In Section~\ref{sec:mc_dpss_blind_fh_reconstruction}, we begin with a discussion of the MC Discrete-Time Equivalent Linear Measurement System (MC-DTLMS). We then show how this framework facilitates segment-based, blind analog FH signal recovery. Moving on, we explain the utility of DPSS vectors as an efficient dictionary for representing sampled FH signals. Accordingly, the use of the DPSS dictionary leads to a reduction in the scale of the signal recovery problem.

In Section~\ref{sec:mc_dpss_simulation}, we use simulations to evaluate the signal reconstruction performance from sub-Nyquist rate MC samples using our segment-based recovery framework. We conclude in Section~\ref{sec:mc_dpss_conclusion} with open questions.

\section{Preliminaries}
\label{sec:mc_dpss_preliminaries}

In this section, after establishing some definitions and notation, we briefly review the fundamental principles of MC sampling and the application of MC sampling in blind multiband signal reconstruction. The interested reader is referred to \cite{544131,868487,950786,pfeng_phd,4601017,5743619,lexa_1} for more details. We also discuss certain CS concepts which will be helpful for subsequent developments in this paper.

\subsection{Definitions and Notation}
\label{sec:mc_dpss_definitions_notations}

The CTFT of a continuous time signal $x(t)$ is denoted by $X(f)$ where $$X(f)=\int_{-\infty}^{\infty}x(t)e^{-j2\pi ft}dt$$ and the discrete time Fourier transform (DTFT) of a discrete time signal $x(k)$ sampled at an interval of $T_s$ seconds/sample is denoted by $\dtft{X}{T_s}$ where $$\dtft{X}{T_s}=\sum_{k=-\infty}^{\infty}x(k)e^{-j2\pi fT_sk}.$$ As a function of $f$, $\dtft{X}{T_s}$ is periodic with period $1/T_s$. Matrices shall be denoted by bold uppercase letters and vectors by bold lowercase letters. For a given matrix $\boldvec{A}\in\mathds{C}^{m\times n}$, $\boldvec{A}^*$ denotes the Hermitian transpose of $\boldvec{A}$. The space spanned by the columns of $\boldvec{A}$ is referred to as the range space of $\boldvec{A}$ and denoted by $\text{range}(\boldvec{A})$.

\subsection{Multi-Coset (MC) Sampling}
\label{sec:mc_dpss_mc_sampling}

\begin{figure}[t]
	\centering
	\begin{minipage}{0.5\textwidth}
		\centering
		\includegraphics[width=0.6\textwidth]{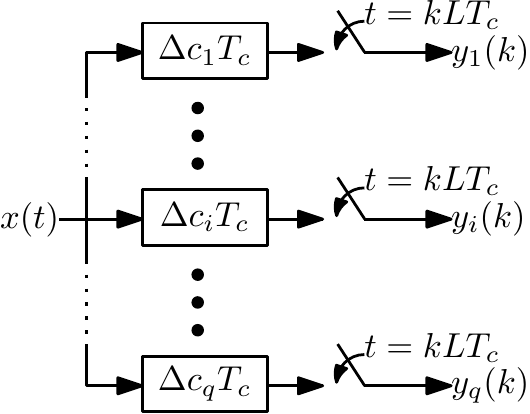}
		\subcaption{A $\mc{T_c}{L}{q}{C}$ MC sampling system.		\label{fig:mc_sampling}}
	\end{minipage}\\
	\begin{minipage}{0.5\textwidth}
		\centering
		\hspace{2ex}
		\includegraphics[width=0.7\textwidth]{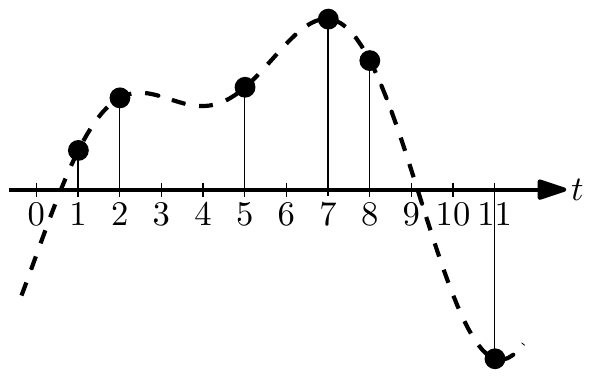}
		\subcaption{An MC sampling pattern with $T_c=1$, $L=6$, $q=3$, and $C=\{1,2,5\}$.\label{fig:mc_sampling_pattern_1}}
	\end{minipage}
		\caption{The MC sampling scheme.}
\end{figure}

MC sampling is a periodic multi-channel nonuniform sampling protocol. We denote an MC sampling system by $\mc{T_{c}}{L}{q}{C}$, where $T_{c}$ is a ``base'' sampling interval that is less than or equal to the input signal's uniform Nyquist sampling interval $T_{\text{nyq}}$, $L>0$ is an integer, $q$ is the number of MC channels, and the set $C$ contains $q$ distinct integers $c_i$ such that $0\le c_i\le L-1$. In the $i$th channel (also referred to as the $i$th active coset), the input signal $x(t)$ is first offset by $c_iT_c$ seconds in time and then sampled uniformly at the interval of $LT_{c}$ seconds. On average, the MC sampler acquires nonuniform samples of the input signal at the rate of $\frac{q}{LT_c}$ samples/second and such samples are referred to as {\em sub-Nyquist} when $\frac{q}{LT_c} < \frac{1}{T_{\text{nyq}}}$. Figures~\ref{fig:mc_sampling} and~\ref{fig:mc_sampling_pattern_1} show the MC sampling system and an example of a MC sampling pattern, respectively.

\begin{figure}[t]
	\begin{minipage}{0.45\textwidth}
		\centering
		\includegraphics[width=0.95\textwidth]{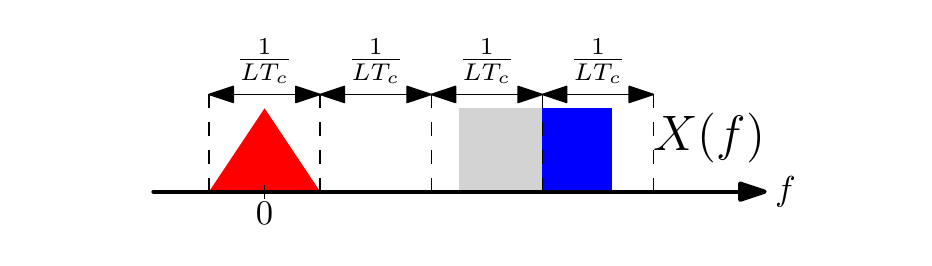}
		\subcaption{{\small Signal spectrum.}\label{fig:signal_spectrum}}
	\end{minipage}\\
	\begin{minipage}{0.45\textwidth}
		\centering
		\includegraphics[width=0.75\textwidth]{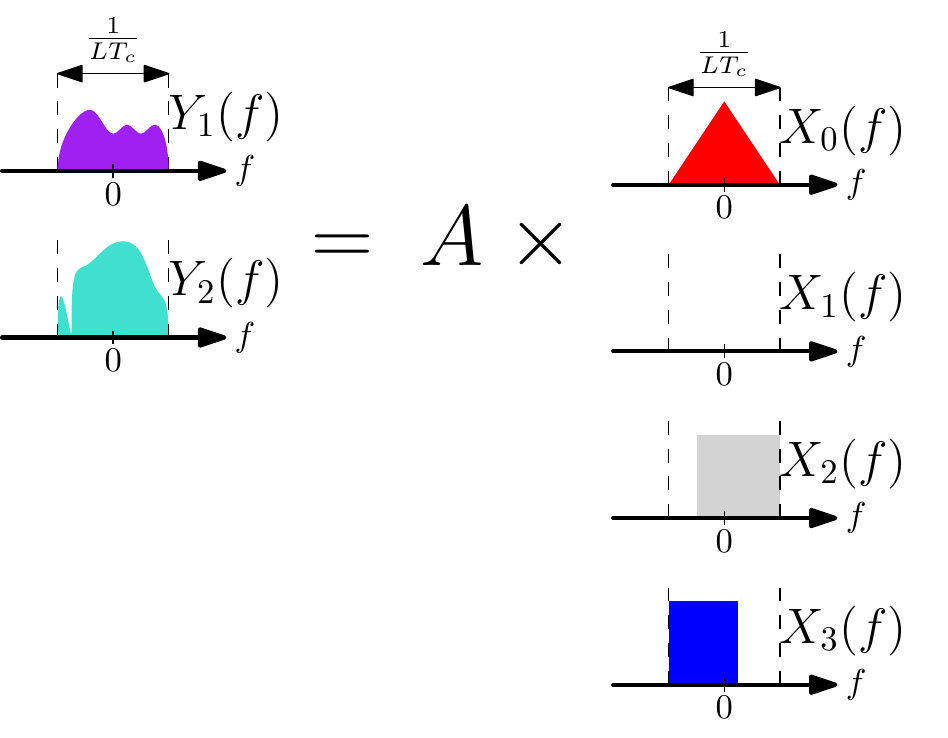}
		\subcaption{{\small The frequency domain linear system of~\eqref{eqn:mc_mat_vec_form} where $Y_i(f)$ is the $i$th entry of $\boldvec{y}(f)$ with $q=2$, $L=4$ and $X_1(f)=0$.}\label{fig:t_f_correspond_f}}
	\end{minipage}
	\caption{\label{fig:t_f_correspond_slicing} Spectral slicing in an MC system.}
\end{figure}

The MC sample stream at the output of the $i$th coset is given by $y_i(k)=x((kL+c_i)T_c)$ and has the corresponding DTFT
\begin{align}
\label{eqn:ith_coset_dtft}
&\dtft{Y_i}{LT_c}=\frac{1}{LT_c}\sum_{l=0}^{L-1} X\left(f+\frac{l}{LT_c}\right)e^{j2\pi c_iT_c\left(f+\frac{l}{LT_c}\right)},\\
&f\in[-1/(2LT_c),1/(2LT_c)],\nonumber
\end{align}
where $X(f)$ denotes the CTFT of the input signal $x(t)$. Rearranging \eqref{eqn:ith_coset_dtft} in matrix-vector form gives the following:
\begin{equation}
\boldvec{y}(f)=\boldvec{A}\boldvec{x}(f),\:f\in\left[-1/(2LT_c),1/(2LT_c)\right]
\label{eqn:mc_mat_vec_form}
\end{equation}
where
\[
\boldvec{y}(f) =LT_c \begin{bmatrix} e^{-j2\pi fc_1T_c}\dtft{Y_1}{LT_c} \\ \vdots \\ e^{-j2\pi fc_qT_c}\dtft{Y_q}{LT_c}\end{bmatrix} \in \mathds{C}^q,
\]
the matrix $\boldvec{A}\in\mathds{C}^{q\times L}$ contains entries
\begin{equation}
\boldvec{A}_{il}=e^{j2\pi c_il/L},
\label{eqn:A_entries}
\end{equation}
\begin{equation*}
\boldvec{x}(f)=\begin{bmatrix}
 X_0(f)&\cdots&X_{L-1}(f)
 \end{bmatrix}^T\in\mathds{C}^L
 %\label{eqn:x_spectrum_vector}
\end{equation*}
with
\begin{equation*}
X_l(f)=X\left(f+\frac{l}{LT_c}\right)\boldvec{1}(f)_{[-1/(2LT_c),1/(2LT_c)]}
%\label{eqn:spectral_slice_L}
\end{equation*}
and
\begin{equation*}
\boldvec{1}(f)_{\mathcal{F}} = \left\{
\begin{array}{lr}
1 & : f\in\mathcal{F}\\
0 & : f \notin\mathcal{F}
\end{array}
\right.
\end{equation*}
denotes the the indicator function of the set $\mathcal{F}$. In words, \eqref{eqn:mc_mat_vec_form} expresses the spectrum of the channel outputs $\boldvec{y}(f)$ (observed) as linear measurements of spectral slices of the input signal $\boldvec{x}(f)$ (unknown) with each spectral slice having a width of $1/(LT_c)$ Hz. In particular, these spectral slices $X_l(f)$ correspond to spectral slices of the input signal shifted to the baseband interval $\left[-1/(2LT_c),1/(2LT_c)\right]$. Notwithstanding the matrix-vector form of \eqref{eqn:mc_mat_vec_form}, this equation holds for all values of $f\in\left[-1/(2LT_c),1/(2LT_c)\right]$ and therefore describes an \emph{infinite} number of linear systems. Figures~\ref{fig:signal_spectrum} and~\ref{fig:t_f_correspond_f} show the input signal spectrum and the corresponding conceptual spectral slicing of the input signal at a frequency resolution of $1/LT_c$, respectively.

In the following sections, we review the necessary conditions for \eqref{eqn:mc_mat_vec_form} to have a unique solution and, in the context of blind spectrum sensing, for the modified MUSIC algorithm proposed previously \cite{544131,pfeng_phd} to solve \eqref{eqn:mc_mat_vec_form} without having to discretize $f$.

\subsection{Blind Multiband Signal Reconstruction From Sub-Nyquist Rate MC Samples}
\label{sec:mc_dpss_blind_multiband_signal_reconstruction_from_mc_samples}

In the context of the blind multiband signal model where only a few spectral bands are occupied, one can expect $\boldvec{x}(f)$ to contain a similar number of nonzero slices. Suppose the index set $\mathcal{I_M}$ contains the $p$ indices of the nonzero slices of $\boldvec{x}(f)$. Then \eqref{eqn:mc_mat_vec_form} can be equivalently reformulated as
\begin{equation}
\boldvec{y}(f)=\boldvec{A}_{\mathcal{I_M}}\boldvec{x}_{\mathcal{I_M}}(f),
\label{eqn:mc_mat_vec_form_reduced}
\end{equation}
where $\boldvec{A}_{\mathcal{I_M}}\in\mathds{C}^{q\times p}$ is the submatrix formed by taking the columns of $\boldvec{A}$ corresponding to the indices in $\mathcal{I_M}$ and $\boldvec{x}_{\mathcal{I_M}}(f)\in\mathds{C}^p$ is the reduced vector containing only the nonzero slices of $\boldvec{x}(f)$. A well-known necessary condition for a unique solution to \eqref{eqn:mc_mat_vec_form_reduced} for any $\boldvec{x}_{\mathcal{I_M}}(f)$ requires that $\boldvec{A}_{\mathcal{I_M}}$ have full column rank for all possible index sets $\mathcal{I_M}$. In turn, this condition requires $\text{spark}(\boldvec{A})=p+1$, where $\text{spark}(\boldvec{A})$ is the minimum number of linearly dependent columns of $\boldvec{A}$. For the matrix $\boldvec{A}$ with entries as given in \eqref{eqn:A_entries}, the interested reader is referred to \cite{4601017} for a detailed discussion on factors affecting $\boldvec{A}$ which in turn affect signal reconstruction quality.

Assuming the necessary condition for a unique solution to \eqref{eqn:mc_mat_vec_form} is satisfied, we now discuss previously proposed blind multiband signal reconstruction techniques in the segment-less reconstruction framework.

When the index set $\mathcal{I_M}$ is known, solving \eqref{eqn:mc_mat_vec_form} can be reduced to solving \eqref{eqn:mc_mat_vec_form_reduced} and its solution is simply given as the following:
\begin{equation}
\boldvec{x}_{\mathcal{I_M}}(f)=\boldvec{A}_{\mathcal{I_M}}^\dagger\boldvec{y}(f),
\label{eqn:pseudoinverse_solve}
\end{equation}
where $\boldvec{A}_{\mathcal{I_M}}^\dagger$ is the pseudoinverse of $\boldvec{A}$. Let us remind the reader that solving \eqref{eqn:pseudoinverse_solve} is equivalent to solving for the nonzero spectral slices of the input signal at spectral resolution $1/(LT_c)$ Hz. In practice, multiband signal reconstruction is performed in the time domain rather than the frequency domain, largely due to the convenience of time domain processing\cite{544131,pfeng_phd}, as discussed in the sequel.

Due to the delays ($c_iT_c$'s) introduced in each channel of the MC sampler and since the sampling rate of each channel is decimated by a factor of $L$, time domain reconstruction begins with interpolation of the MC sample outputs to a sampling rate of $1/T_c$ and then involves correcting for the delays. This can be easily achieved by computing
\begin{equation}
z_i(k)=y_i\left(\frac{k-c_i}{L}\right).
\label{eqn:interpolate_offset_correct_mc_output}
\end{equation}
Subsequently, the interpolated (and delay corrected) MC time samples ($z_i(k)$'s) are linearly combined using $\boldvec{A}^\dagger_{\mathcal{I_M}}$ to obtain the time domain sequences of the input signal corresponding to the baseband spectral slices $X_l(f)$. Finally, each of these time domain sequences are frequency shifted to their original corresponding spectral slice location and summed to obtain a reconstructed sampled version $x(k)$ of the input signal $x(t)$. In short, time domain signal reconstruction recovers $x(k)$ by computing
\begin{equation*}
x(k)=\frac{1}{LT_c}\sum_{i=1}^{q}\sum_{\ell=1}^{|\mathcal{I_M}|}\beta_{\ell i}z_i(k)\exp\left(\frac{j2\pi\ell k}{L}\right),
\end{equation*}
where $\beta_{\ell i}$ correspond to the entries of $\boldvec{A}_{\mathcal{I_M}}^\dagger$.

When the index set $\mathcal{I_M}$ is unknown, an additional step is required to first obtain $\mathcal{I_M}$ before the above steps can be applied. This additional step is referred to as the modified MUSIC algorithm in \cite{544131,pfeng_phd,4601017}. First, consider the covariance matrix of $\boldvec{y}(f)$,
\begin{align}
\boldvec{R_y}&=\int_{f\in [-1/(2LT_c),1/(2LT_c)]}\boldvec{y}(f)\boldvec{y}(f)^*df \nonumber \\
&=\boldvec{A}\left[\int_{f\in[-1/(2LT_c),1/(2LT_c)]}\boldvec{x}(f)\boldvec{x}(f)^*df\right]\boldvec{A}^*\nonumber\\
&=\boldvec{AR_xA}^*,
\label{eqn:covariance_mat_y}
\end{align}
where $\boldvec{R_x}$ denotes the covariance matrix of $\boldvec{x}(f)$. %Given the multiband signal model, only $|\mathcal{I_M}|$ spectral slices will be active and hence only $|\mathcal{I_M}|$ entries in $\boldvec{x}(f)$ are nonzero.
Using \eqref{eqn:mc_mat_vec_form_reduced}, \eqref{eqn:covariance_mat_y} can be reexpressed as
\begin{equation}
\boldvec{R_y}=\boldvec{A}_{\mathcal{I_M}}\boldvec{R}_{\boldvec{x}_{\mathcal{I_M}}}\boldvec{A}^*_{\mathcal{I_M}},
\label{eqn:covariance_mat_y_1}
\end{equation}
where $\boldvec{R}_{\boldvec{x}_{\mathcal{I_M}}}$ is the covariance matrix of $\boldvec{x}_{\mathcal{I_M}}$. In terms of its eigen-decomposition, $\boldvec{R_y}$ can be expressed as $\boldvec{R_y}=\boldvec{U\Lambda U}^*$ where $\boldvec{U}$ contains the eigenvectors of $\boldvec{R_y}$ and $\boldvec{\Lambda}$ is a diagonal matrix containing the eigenvalues of $\boldvec{R_y}$. If $\text{rank}(\boldvec{R}_{\boldvec{x}_{\mathcal{I_M}}})=|\mathcal{I_M}|=p$ (i.e., full rank), then $\text{rank}(\boldvec{R_y})=p$ since $\text{rank}(\boldvec{A}_{\mathcal{I_M}})=p$ as necessitated by the unique solution condition. (The interested reader is referred to~\cite{4601017,lexa_1} for more details on this argument.) As such $\boldvec{R_y}$ will have $p$ nonzero eigenvalues and correspondingly,
\begin{equation}
\boldvec{R_y}=\boldvec{U_x\Lambda_xU_x^*}+\boldvec{U_n\Lambda_nU_n^*}=\boldvec{U_x\Lambda_xU_x^*},
\label{eqn:covariance_mat_y_eig}
\end{equation}
where $\boldvec{U_x}$ contains the eigenvectors corresponding to the $p$ nonzero eigenvalues, $\boldvec{\Lambda_x}$ is a diagonal matrix containing the nonzero eigenvalues, $\boldvec{U_n}$ contains the eigenvectors corresponding to the zero eigenvalues of $\boldvec{R_y}$ and $\boldvec{\Lambda_n}$ is a matrix containing zero entries of appropriate dimensions. Eqn.~\eqref{eqn:covariance_mat_y_eig} implies that $\text{range}(\boldvec{U_x})=\text{range}(\boldvec{R_y})$ which, in turn, implies that $\text{range}(\boldvec{U_x})=\text{range}(\boldvec{A}_{\mathcal{I_M}})$ due to \eqref{eqn:covariance_mat_y_1}. On the other hand, $\text{range}(\boldvec{U_n})$ is orthogonal to $\text{range}(\boldvec{U_x})$ and hence is also orthogonal to $\text{range}(\boldvec{A}_{\mathcal{I_M}})$. Therefore one can obtain $\mathcal{I_M}$ by projecting all columns of $\boldvec{A}$ onto $\boldvec{U_n}$ (by computing $\boldvec{U^*_n}\boldvec{A}$) and constructing $\mathcal{I_M}$ using indices corresponding to the columns that give the $p$ smallest projections in terms of the $\ell_2$-norm. We note that in order to estimate $\boldvec{U_n}$, a necessary and sufficient condition is that $q\ge p+1$ so that there exists at least one eigenvector to represent the space spanned by $\boldvec{U_n}$. Correspondingly, this condition also translates to a \emph{deterministic} guarantee for perfect multiband signal recovery when $\boldvec{R_x}$ is full rank. In the worst case scenario when $\text{rank}(\boldvec{R_x})=1$, this guarantee becomes (the less favorable) $q\ge 2p$. When $\boldvec{R_x}$ is full rank, the deterministic guarantee results in a minimum average sampling rate of $(p+1)/(LT_c)$. Letting $\Omega_L$ denote the spectral occupancy at resolution $L$, we have $\Omega_L=p/L$. In this setting, the modified MUSIC algorithm is said to approach the Landau rate, which equals $\Omega/T_c$, where $\Omega$ is the spectral occupancy, \emph{asymptotically} for almost all multiband signals since $\Omega_L\to\Omega$ as $L\to\infty$. After $\mathcal{I_M}$ has been identified for a multiband signal, one only needs to solve \eqref{eqn:pseudoinverse_solve} for subsequent MC samples since the spectral support does not change.

\subsection{Finite-Dimensional CS Reconstruction Framework}
\label{sec:mc_dpss_cs_prelim}

The emerging theory of CS has enabled the efficient acquisition of sparse or compressible signals via low-rate random projections. In its classical form, the CS framework deals with a linear system of under-determined equations
\begin{equation}
\boldvec{y}=\boldvec{Ax},
\label{eqn:cs_smv}
\end{equation}
where the measurements $\boldvec{y}\in\mathds{C}^{q}$, sensing matrix $\boldvec{A}\in\mathds{C}^{q\times L}$, unknown signal vector $\boldvec{x}\in\mathds{C}^L$ and $q\ll L$. Under certain conditions it may be possible to recover $\boldvec{x}$ from $\boldvec{y}$ (even though $q\ll L$), for example if $||\boldvec{x}||_0\ll L$, where $||\boldvec{x}||_0$ denotes the number of nonzero entries of $\boldvec{x}$. A signal $\boldvec{x}$ is said to be $p$-sparse if $||\boldvec{x}||_0=p$ and compressible if it has $p$ significant entries (with the rest of the entries being small).

The matrix $\boldvec{A}$ satisfies the {\em Restricted Isometry Property} (RIP) of order $p$ with isometry constant $\delta_p\in (0,1)$ if
\begin{equation*}
(1-\delta_p)||\boldvec{x}||_2\le ||\boldvec{Ax}||_2\le (1+\delta_p)||\boldvec{x}||_2
\end{equation*}
holds for all $p$-sparse vectors $\boldvec{x}$. It was shown previously \cite{Candes06} that exact recovery of $\boldvec{x}$ from $\boldvec{y}$ can be achieved by solving a convex optimization problem (specifically, $\ell_1$ minimization) provided $\boldvec{A}$ satisfies the RIP of order $2p$ with $\delta_{2p}$ small. Further, it was shown in \cite{Rudelson06} that one can obtain such an $\boldvec{A}$ (with entries as defined in \eqref{eqn:A_entries}) by randomly selecting rows of the DFT matrix, with high probability if $q=\mathcal{O}(p\log^4(L))$. In addition, numerous greedy iterative recovery algorithms such as OMP \cite{4385788} and CoSaMP \cite{Needell10} have also been proposed. We note that the cited recovery algorithms have been shown to be robust to noise and return an approximate solution when $\boldvec{x}$ is compressible. The linear system of \eqref{eqn:cs_smv} is also referred to the single measurement vector (SMV) problem in the CS literature.

As an extension to the SMV linear system of \eqref{eqn:cs_smv}, multiple measurement vector (MMV) linear systems have also been considered \cite{Tropp_2006_1,4014378,Tropp_2006_2,6145474}. In particular, an MMV linear system of under-determined equations has the following form:
\begin{equation}
\boldvec{Y}=\boldvec{AX},
\label{eqn:cs_mmv}
\end{equation}
where $\boldvec{Y}\in\mathds{C}^{q\times r}$, $\boldvec{A}\in\mathds{C}^{q\times L}$ and $\boldvec{X}\in\mathds{C}^{L\times r}$ with an assumed structure in $\boldvec{X}$. The assumed structure is typically in the form of joint sparsity across the columns of $\boldvec{X}$ such that the index set containing the locations of the nonzero entries for every column of $\boldvec{X}$ is identical. For the MMV linear system, numerous recovery algorithms such as S-OMP \cite{Tropp_2006_1,4014378}, Mixed Norm~\cite{Tropp_2006_2}, and the previously cited modified MUSIC algorithm \cite{544131,pfeng_phd} have been proposed.

It has been shown in numerous works~\cite{pfeng_phd, 4014378,4601017,6145474} that guaranteeing a unique solution to the MMV linear system of \eqref{eqn:cs_mmv} requires
\begin{equation*}
	||\boldvec{X}||_0\le\frac{\text{spark}(\boldvec{A})+\text{rank}(\boldvec{Y})-1}{2},
%\label{eqn:mmv_unique_condition}
\end{equation*}
where $||\boldvec{X}||_0$ denotes the number of nonzero rows of $\boldvec{X}$. While it has been suggested that the ability to recover the true support of $\boldvec{X}$ improves as the number $r$ of measurement vectors increases (the number of columns of $\boldvec{Y}$), \eqref{eqn:cs_mmv} reveals the important dependence of the recovery performance on $\text{rank}(\boldvec{Y})$. Intuitively, one cannot expect to increase the ability to recover the true support of $\boldvec{X}$ when the additional columns of $\boldvec{Y}$ do not providing more information about the true support of $\boldvec{X}$, which is the case when an increase in $r$ does not commensurate with an increase in $\text{rank}(\boldvec{Y})$. As we discuss in the sequel, this bound on the maximum number of nonzero rows of $\boldvec{X}$ can be used to establish a bound on the blind recovery of FH signals when the CS MMV framework is used for recovery.

\subsection{Discrete Prolate Spheroidal Sequences (DPSSs)}
\label{sec:mc_dpss}

In a series of seminal papers, Slepian et al. \cite{slepian_1,slepian_2} and Landau et al. \cite{landau_2,landau_3} derived and examined properties of the Prolate Spheroidal Wave Functions (PSWFs) which have important implications in time-frequency analysis. Slepian later examined discrete versions of the PSWFs named Discrete Prolate Spheroidal Sequences (DPSSs) in \cite{slepian_3}. In short, PSWFs and correspondingly, DPSSs have maximal energy concentration given a time interval (in which they are highly concentrated) and frequency interval (in which they are strictly bandlimited). The PSWFs and DPSSs are defined to be the eigenfunctions and eigenvectors, respectively of a two step procedure which first time-limits the function (sequence) and then band-limits it.

The DPSSs are parameterized by an integer $N_D\in\mathds{Z}^+$ and $W_D$, where $0\le W_D \le\frac{1}{2}$. Specifically, the DPSSs are a collection of $N_D$ discrete time sequences that are strictly bandlimited to the digital frequency $|f|\le W_Df_s$ where $f_s$ is the sampling frequency, and highly concentrated in the time index range $\{0,1,\dots,N_D-1\}$. Let $\mathcal{B}_{W_D}$ denote an operator that takes as input a discrete time signal, bandlimits its DTFT to the frequencies $|f|\le W_Df_s$, and then returns the resulting discrete time signal. Next, let $\mathcal{T}_{N_D}$ denote an operator that takes as input an infinite length discrete time signal and returns an infinite length discrete time signal with all entries outside the index range $\{0,1,\dots,N_D-1\}$ set to zero. Formally, DPSSs are defined as follows.
\begin{definition}
	\cite{slepian_3} Given $N_D$ and $W_D$, the Discrete Prolate Spheroidal Sequences \textnormal{(DPSSs)} are a collection of $N_D$ real-valued discrete-time sequences ${s}^{(0)}_{N_D,W_D},{s}^{(1)}_{N_D,W_D},\dots,{s}^{(N_D-1)}_{N_D,W_D}$ that, along with the corresponding scalar eigenvalues $$1>\lambda^{(0)}_{N_D,W_D}>\lambda^{(1)}_{N_D,W_D}>\cdots>\lambda^{(N_D-1)}_{N_D,W_D}>0,$$ satisfy
	\begin{equation*}
	\mathcal{B}_{W_D}(\mathcal{T}_{N_D}(s^{(\ell)}_{N_D,W_D}))=\lambda^{(\ell)}_{N_D,W_D}s^{(\ell)}_{N_D,W_D}
	\end{equation*}
	for all $\ell\in\{0,1,\dots,N_D-1\}$. The DPSSs are normalized so that
	\begin{equation*}
	||\mathcal{T}_{N_D}(s^{(\ell)}_{N_D,W_D})||_2=1
	%\label{eqn:dpss_unit_norm}
	\end{equation*}
	for all $\ell\in\{0,1,\dots,N_D-1\}$.	
	\label{def:dpss}
\end{definition}
The DPSSs are orthogonal both on $\{0,1,\dots,N_D-1\}$ and on $\mathds{Z}$. In this work, we are primarily interested in truncated time-limited DPSSs, referred to as {\em DPSS vectors} and denoted by $\mathcal{D}(N_D,W_D)$.

\begin{definition}
	\cite{davenport2012compressive} Given $N_D$ and $W_D$, the \textnormal{DPSS vectors} $\boldvec{s}^{(0)}_{N_D,W_D},\boldvec{s}^{(1)}_{N_D,W_D},\dots,\boldvec{s}^{(N_D-1)}_{N_D,W_D}\in\mathds{R}^{N_D}$ are defined by restricting the time-limited DPSSs to the index range $n=0,1,\dots,N_D-1$:
	\begin{equation}
	\boldvec{s}^{(\ell)}_{N_D,W_D}[n]:=\mathcal{T}_{N_D}(s^{(\ell)}_{N_D,W_D})[n]=s^{(\ell)}_{N_D,W_D}[n]
	\label{eqn:dpss_vector}
	\end{equation}
	for all $\ell,n\in\{0,1,\dots,N_D-1\}$.
	\label{def:dpss_vector}
\end{definition}

It follows that $\mathcal{D}(N_D,W_D)$ form an orthonormal basis for $\mathds{C}^{N_D}$ (or for $\mathds{R}^{N_D}$). A unique characteristic of DPSSs is that the first $2N_DW_D$ eigenvalues cluster close to $1$ while the remaining eigenvalues cluster close to $0$. Formally, this characteristic is captured in the following lemmas.
\begin{lemma}
	\textnormal{(Eigenvalues that cluster near one \cite{slepian_3}).} Suppose that $W_D$ is fixed, and let $\epsilon\in(0,1)$ be fixed. Then there exist constants $C_1$, $C_2$ (where $C_2$ may depend on $W_D$ and $\epsilon$) and an integer $N_0$ (which may also depend on $W_D$ and $\epsilon$) such that
	\begin{equation*}
	\lambda^{(\ell)}_{N_D,W_D}\ge 1-C_1e^{-C_2N_D},
	%\label{eqn:dpss_eigenvalue_one}
	\end{equation*}
	for all $\ell\le 2N_DW_D(1-\epsilon)$ and all $N_D\ge N_0$.
	\label{lm:dpss_eigenvalue_one}
\end{lemma}
\begin{lemma}
	\textnormal{(Eigenvalues that cluster near zero \cite{slepian_3}).} Suppose that $W_D$ is fixed, and let $\epsilon\in(0,\frac{1}{2W_D}-1)$ be fixed. Then there exist constants $C_3$, $C_4$ (where $C_4$ may depend on $W_D$ and $\epsilon$) and an integer $N_1$ (which may also depend on $W_D$ and $\epsilon$) such that
	\begin{equation*}
	\lambda^{(\ell)}_{N_D,W_D}\le C_3e^{-C_4N_D},
	%\label{eqn:dpss_eigenvalue_zero_1}
	\end{equation*}
	for all $\ell\ge 2N_DW_D(1+\epsilon)$ and all $N_D\ge N_1$. Alternatively, suppose that $W_D$ is fixed, and let $\alpha>0$ be fixed. Then there exist constants $C_5$, $C_6$ and an integer $N_2$ (where $N_2$ may depend on $W_D$ and $\alpha$) such that
	\begin{equation*}
	\lambda^{(\ell)}_{N_D,W_D}\le C_5e^{-C_6N_D},
	%\label{eqn:dpss_eigenvalue_zero_2}
	\end{equation*}
	for all $\ell\ge 2N_DW_D+\alpha\log(N_D)$ and all $N_D\ge N_2$.
	\label{lm:dpss_eigenvalue_zero}
\end{lemma}

As discussed further in Section~\ref{sec:mc_segment_recovery_dpss}, we will be interested in using DPSS vectors to build a dictionary for representing sampled signal vectors. Lemmas~\ref{lm:dpss_eigenvalue_one} and~\ref{lm:dpss_eigenvalue_zero} imply that only the first $\approx 2N_DW_D$ DPSS vectors are required to capture the energy of length-$N_D$ signal vectors that arise from time-limiting discrete-time signals that are bandlimited to frequencies $|f|\le W_Df_s$. The following theorem from \cite{davenport2012compressive} gives an approximation guarantee for the representation of discrete-time, approximately bandlimited, truncated time-limited signals using slightly more than the first $2N_DW_D$ DPSS vectors.
\begin{thm}
	\textnormal{\cite{davenport2012compressive}} Let $x(k)=\mathcal{T}_{N_D}(x(k))$ be a time-limited sequence, and suppose that $x(k)$ is approximately bandlimited to the frequency range $f\in[-W_Df_s,W_Df_s]$ such that for some $\delta$,
	\begin{equation*}
	||\mathcal{B}_{W_D}x||^2_2\ge (1-\delta)||x||^2_2.
	\end{equation*}
    Let $\boldvec{x}\in\mathds{C}^{N_D}$ denote the vector formed by restricting $x(k)$ to the indices $k=0,1,\dots,N_D-1$. Set $k_D=2N_DW_D(1+\epsilon)$ and let
    \[
    \boldvec{Q}=[\boldvec{s}^{(0)}_{N_D,W_D} ~ \boldvec{s}^{(1)}_{N_D,W_D} ~ \cdots ~ \boldvec{s}^{(k_D-1)}_{N_D,W_D}].
    \]
    Then for $N_D\ge N_1$,
	\begin{equation*}
	||\boldvec{x}-\boldvec{P_Qx}||^2_2\le (\delta+N_DC_3e^{-C_4N_D})||\boldvec{x}||^2_2,
	%\label{eqn:dpss_approx_quality_2}
	\end{equation*}
	where $\boldvec{P_Qx}$ denotes the orthogonal projection of $\boldvec{x}$ onto the space spanned by the columns of $\boldvec{Q}$, and $C_3$, $C_4$, and $N_1$ are as specified in Lemma~\ref{lm:dpss_eigenvalue_zero}.
	\label{th:dpss_approx_quality_2}
\end{thm}

\section{Blind Analog FH Signal Reconstruction from MC Samples}
\label{sec:mc_dpss_blind_fh_reconstruction}

We are now ready to answer the central question of this paper: How can one reconstruct signals belonging to $\afh{N}{B}{T}$ from sub-Nyquist rate MC samples? Unfortunately, it would be difficult to apply \eqref{eqn:mc_mat_vec_form} directly in reconstructing an FH signal. In particular, in order to obtain the term on the left-hand side of this equation, $\boldvec{y}(f)$, it may be necessary to observe the channel outputs ($y_i(k)$'s) over a long time duration so that an accurate estimate of $\dtft{Y_i}{LT_c}$ can be formed. As we have explained, while such long sensing intervals may be feasible for multiband signals, they may be infeasible for FH signals where the spectral support changes (and thus, effectively increases) over time. Fortuitously, as we will show, this difficulty can be avoided by converting the infinite linear system of \eqref{eqn:mc_mat_vec_form} into an equivalent form in the discrete-time domain. We refer to this equivalent form as the MC Discrete-Time Equivalent Linear Measurement System (MC-DTLMS).

\subsection{MC Discrete-Time Equivalent Linear Measurement System (MC-DTLMS)}
\label{sec:mc_dt_equivalent_linear_measurement_sys}

Our goal now is to highlight a salient aspect of the versatility of the MC sampling scheme which has not been emphasized in previous literature focusing on multiband signal reconstruction. The MC-DTLMS is summarized in the following lemma.

\begin{lemma}
	The infinite linear system of \eqref{eqn:mc_mat_vec_form} is equivalent to the system of equations
	\begin{equation}
	\boldvec{z}(k)=\boldvec{Ax}_{\text{bb}}(k)\;\;\forall k,
	\label{eqn:mc_dt_mat_vec_form}
	\end{equation}
	where
	\begin{equation*}
	\boldvec{z}(k)=
	\begin{bmatrix}
	z_1(k)&\cdots&z_q(k)
	\end{bmatrix}^T,
	\end{equation*}
	$z_i(k)$ is as given in \eqref{eqn:interpolate_offset_correct_mc_output},
	$\boldvec{A}$ is as given in \eqref{eqn:A_entries},
	\begin{equation*}
	\boldvec{x}_{\text{bb}}(k)=
	\begin{bmatrix}
	x_{0,\text{bb}}(k)&\cdots&x_{L-1,\text{bb}}(k)
	\end{bmatrix}^T,
	%\label{eqn:x_dt_vector}
	\end{equation*}
	where
		\begin{equation}
		\dtft{X_{\ell,\text{bb}}}{T_c}\triangleq
		\begin{cases}
		X_\ell(f), & f\in\mathcal{F}_0, \\
		0,       & \text{otherwise,}
		\end{cases}
        \label{eqn:timedomainequivDTFTxbb}
		\end{equation}
    is the DTFT of the $\ell$th sample stream $x_{\ell,\text{bb}}(k)$ and $\mathcal{F}_0=[-1/(2LT_c),1/(2LT_c)]$.
	\label{lm:mc_dt_mat_vec_form}
\end{lemma}
\textbf{Proof}: See Appendix~\ref{adx:mc_dt_mat_vec_form}.

This straightforward but important lemma relates the interpolated, offset corrected MC samples $z_i(k)$ to the sample streams of baseband signals $x_{\ell,\text{bb}}(k)$ having DTFTs identical to the spectral slices $X_\ell(f)$ at a sampling rate of $1/T_c$. As our primary interest is the reconstruction of Nyquist rate samples of the input signal $x(t)$, the use of Lemma~\ref{lm:mc_dt_mat_vec_form} in the sequel is appropriate since $T_c\le T_{\text{nyq}}$.

While previous works on multiband signal reconstruction from sub-Nyquist MC samples have focused on the linear system of \eqref{eqn:mc_mat_vec_form}, Lemma~\ref{lm:mc_dt_mat_vec_form} reveals a salient aspect of the MC sampler which deserves attention in this setting as summarized in the following proposition.
\begin{prop}
	The index set containing the indices of the nonzero slices of $\boldvec{x}(f)$ in the frequency domain linear system of \eqref{eqn:mc_mat_vec_form} is identical to the index set containing the indices of the nonzero sample streams of $\boldvec{x}_{\text{bb}}(k)$ in the discrete-time domain linear system of \eqref{eqn:mc_dt_mat_vec_form}. 
	\label{prop:same_index_set_in_time_freq}
\end{prop}

\begin{figure}[t]
    \centering
    \includegraphics[width=3in]{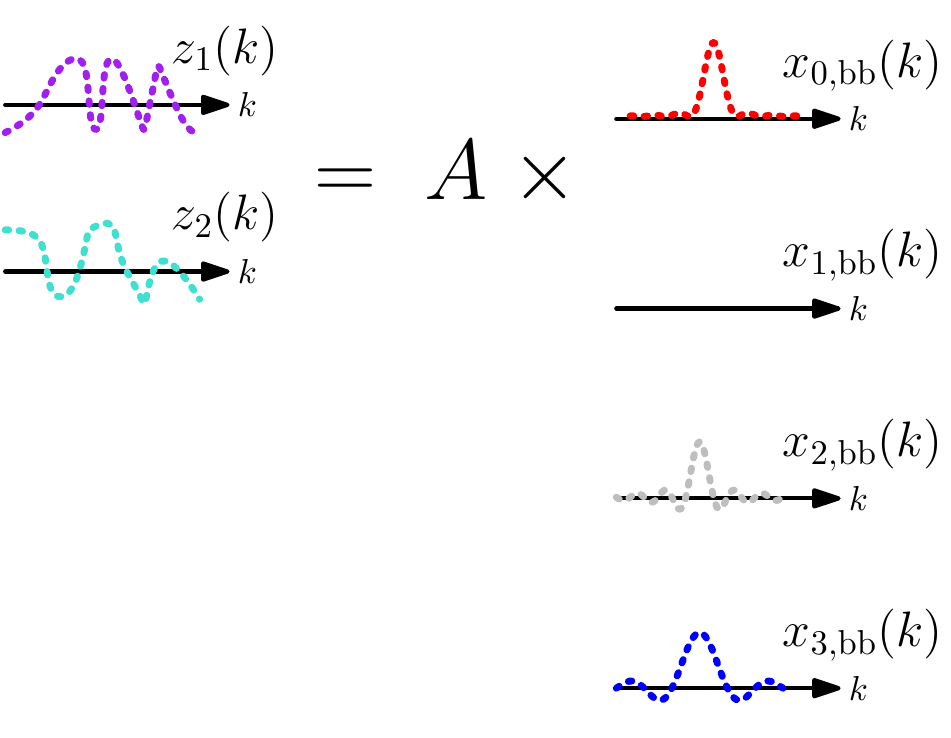}
	\caption{\label{fig:t_f_correspond}An illustration of Proposition~\ref{prop:same_index_set_in_time_freq}: the discrete-time equivalent linear system of~\eqref{eqn:mc_dt_mat_vec_form} corresponding to Figure~\ref{fig:t_f_correspond_f}.}
\end{figure}

Essentially, Proposition~\ref{prop:same_index_set_in_time_freq} states that there exists a direct correspondence between the slices of $\boldvec{x}(f)$ and the sample streams of $\boldvec{x}_{\text{bb}}(k)$ even though the former corresponds to a vector formulation in the frequency domain while the latter corresponds to that in the time domain. Accordingly, an all-zero spectral slice of $\boldvec{x}(f)$ at index $\ell$ corresponds to an all-zero sample stream in $\boldvec{x}_{\text{bb}}(k)$ at the same index $\ell$. Figure~\ref{fig:t_f_correspond} shows an illustration of Proposition~\ref{prop:same_index_set_in_time_freq}; compare with Figure~\ref{fig:t_f_correspond_f}.

The difference between \eqref{eqn:mc_mat_vec_form} and \eqref{eqn:mc_dt_mat_vec_form} is not particularly significant when sampling and reconstructing a multiband signal. In particular, in the case of a multiband signal model, one would expect that, among the $L$ spectral slices that comprise $\boldvec{x}(f)$, few are nonzero. Similarly, among the $L$ infinite sequences that comprise $\boldvec{x}_{\text{bb}}(k)$, few of the sample streams $x_{\ell,\text{bb}}(k)$ are nonzero. Reconstruction could be performed in either domain.

The difference between \eqref{eqn:mc_mat_vec_form} and \eqref{eqn:mc_dt_mat_vec_form} {\em is} significant when sampling and reconstructing an FH signal. In particular, in the case of an FH signal model where samples are collected over a long time duration, many slices of $\boldvec{x}(f)$ will be nonzero as we have explained previously. Consequently, many of the sample steams $x_{\ell,\text{bb}}(k)$ will also be nonzero. Specifically, what this means is that for most values of $\ell$ there will exist some $k$ such that $x_{\ell,\text{bb}}(k) \neq 0$. However, the MC-DTLMS in \eqref{eqn:mc_dt_mat_vec_form} reveals a time-varying structure to sample steams $x_{\ell,\text{bb}}(k)$ that is not apparent from \eqref{eqn:mc_mat_vec_form}. In particular, for any time index $k$, the number of indices $\ell$ for which $x_{\ell,\text{bb}}(k) \neq 0$ should be commensurate with the number of FH radios $N$. This suggests that for any $k$, it may be possible to solve the linear system of \eqref{eqn:mc_dt_mat_vec_form} by exploiting the sparsity among the sample streams. Solving this system repeatedly for every single value of $k$ may be expensive; therefore, we propose to solve this system over time segments of duration commensurate with the minimum HRI $T$. All of this differs from previous works on multiband signal reconstruction from MC samples as discussed previously. We discuss pertinent aspects of solving \eqref{eqn:mc_dt_mat_vec_form} in the following section.

\subsection{An MC Segment-Based Recovery Framework}
\label{sec:mc_dpss_blind_fh_reconstruction_mc_time_domain_equivalent_linear}

Formally, we set up our MC segment-based signal reconstruction problem as follows:
\begin{equation}
\boldvec{Z}=\boldvec{AX}_{\text{bb}},
\label{eqn:finite_mc_dt_mat_vec_form}
\end{equation}
where $\boldvec{Z}\in\mathds{C}^{q\times r}$ with entry $[Z]_{ij}=z_i(k_j)$ in its $i$th row and $j$th column, $\boldvec{A}$ is as given in \eqref{eqn:A_entries}, and $\boldvec{X}_{\text{bb}}\in\mathds{C}^{L\times r}$ with entry $[X_{\text{bb}}]_{ij}=x_{i-1,\text{bb}}(k_j)$ in its $i$th row and $j$th column. The linear system of \eqref{eqn:finite_mc_dt_mat_vec_form} is a restriction of \eqref{eqn:mc_dt_mat_vec_form} to a finite dimension (or segment size) $r$ since the sample streams of $\boldvec{x}_\text{bb}(k)$ correspond to infinite sequences. Accordingly, solving \eqref{eqn:finite_mc_dt_mat_vec_form} repeatedly over consecutive segments of size $r$ recovers consecutive segments of baseband sequences $\boldvec{X}_{\text{bb}}$. In order to reconstruct sampled sequences of the input signal $x(t)$ at rate $1/T_c$, it is necessary to frequency shift the baseband sequences to their corresponding spectral slice location and then sum the frequency-shifted sequences. We note that when the segment size $r=1$, \eqref{eqn:finite_mc_dt_mat_vec_form} turns into the classical SMV problem in CS. When $r>1$, we have an MMV problem, and the structure of $\boldvec{X}_{\text{bb}}$ deserves special mention with regards to the $\afh{N}{B}{T}$ signal model.

\begin{figure}[t]
	\centering
	\includegraphics[width=0.4\textwidth]{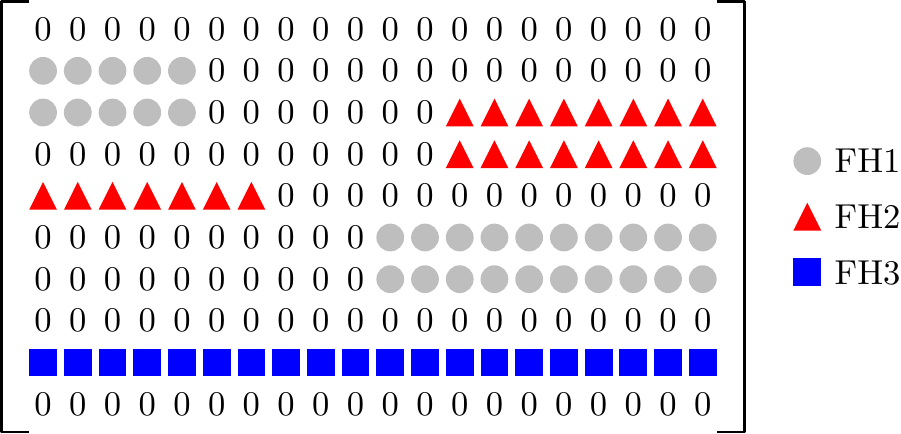}
	\caption{\label{fig:X_bb_matrix}A toy example of the type of structure in $\boldvec{X}_{\text{bb}}\in\mathds{C}^{10\times 20}$ for $N=3$ (denoted by FH1, FH2 and FH3) with identical HRIs, $B\le 1/(LT_c)$, segment size $r<\text{round}(T/T_c)$, as well as the locations of their corresponding nonzero entries in $\boldvec{X}_{\text{bb}}$.}
\end{figure}

Fig.~\ref{fig:X_bb_matrix} shows a toy example of the type of structure that exists in $\boldvec{X}_{\text{bb}}$ when $B\le 1/(LT_c)$ and $r<\text{round}(T/T_c)$; that is, the segment size has a smaller duration than the HRI of the frequency hoppers. We note that setting $L$ such that $B\le 1/(LT_c)$ is a commonly used spectral width configuration in previous \cite{4601017,4749297,544131,868487} multiband signal reconstruction literature. We will make this assumption throughout the rest of this paper. We are primarily interested in the setting where $r<\text{round}(T/T_c)$, since the spectral support ``size'' of the $\afh{N}{B}{T}$ signal is minimal in the sense that each hopper may change frequency at most once.

Due to the definition of $B$ ($\le 1/(LT_c)$) in~\eqref{def:afh_B}, we note that every hop transmitted results in at most 2 nonzero rows of entries over the columns (duration) for which the hop is present. There exist instances when there is only 1 nonzero row of entries if the range of active frequencies corresponding to that particular hop falls within the boundaries of the spectral slice $X_{\ell}(f)$ for the $\ell$th row. Accordingly, for segments when the $N$ hoppers do not switch frequencies, one can expect at most $2N$ nonzero rows in $\boldvec{X}_{\text{bb}}$. On the other hand, for segments when the $N$ hoppers do switch frequencies, one can expect at most $4N$ nonzero rows\footnote{A row is declared nonzero it contains at least one nonzero entry.} in $\boldvec{X}_{\text{bb}}$. The following corollary provides the necessary condition for the existence of a unique solution to \eqref{eqn:finite_mc_dt_mat_vec_form} for a given $\afh{N}{B}{T}$ signal.
\begin{cor}
	(Uniqueness Condition for MC Segment-based Recovery Framework) Let $x(t)\in\afh{N}{B}{T}$. A unique solution exists to \eqref{eqn:finite_mc_dt_mat_vec_form} if
	\begin{equation}
	q>8N-\text{rank}(\boldvec{Z}),
	\label{eqn:min_q_for_unique_sol}
	\end{equation}
	for each segment of size $r$ where $1\le r <\text{round}(T/T_c)$, provided $\text{spark}(\boldvec{A})=q+1$.
	\label{lm:finite_mc_dt_unique_sol}
\end{cor}
\textbf{Proof}: See Appendix~\ref{adx:finite_mc_dt_unique_sol}.

Corollary~\ref{lm:finite_mc_dt_unique_sol} provides a worst case guarantee for obtaining a unique solution to the linear system of \eqref{eqn:finite_mc_dt_mat_vec_form} and hence can be interpreted as a minimal sampling rate requirement. We note that while the $8N$ term of Corollary~\ref{lm:finite_mc_dt_unique_sol} seems excessive, it is due to a worst case consideration. On average, if $T=rnT_c$, where $n\in\mathds{Z}^+$ with $n\gg 1$, one can expect most of the segments to contain hopping signals where the hoppers do not switch frequencies thereby reducing the $8N$ term to $4N$ for such segments. Let us remind the reader that when $r\gg\text{round}(T/T_c)$, the $8N$ term in \eqref{eqn:min_q_for_unique_sol} increases due to the growing spectral support of the $\afh{N}{B}{T}$ signal, which in turn increases the minimum average MC sampling rate required. On the other hand, one can possibly reduce the minimum average MC sampling rate (by reducing $q$) using a $\boldvec{Z}$ with a larger rank, which is often achievable by increasing the segment size $r$. Due to the opposing effects on \eqref{eqn:min_q_for_unique_sol} when increasing $r$, the appropriate choice of $r$ to use translates to a design trade-off that one must consider for a given $\afh{N}{B}{T}$ signal. We shall discuss this design aspect of MC segment-based signal recovery further in Section~\ref{sec:mc_dpss_simulation}.

\subsection{An Efficient Dictionary for the MC Segment-based Recovery Framework}
\label{sec:mc_segment_recovery_dpss}

We now discuss the utility of DPSS vectors as an efficient dictionary for reducing the scale of the recovery problem. Recall that the finite linear system of \eqref{eqn:finite_mc_dt_mat_vec_form} requires solving for the $r$ columns of unknowns in $\boldvec{X}_{\text{bb}}$ given that $\boldvec{Z}\in\mathds{C}^{q\times r}$ and $\boldvec{A}\in\mathds{C}^{q\times L}$. As we discuss in Section~\ref{sec:mc_dpss_simulation}, when solving \eqref{eqn:finite_mc_dt_mat_vec_form}, the resulting solver latency can be high when $r$ is large. However, note that the each row of $\boldvec{Z}$ corresponds to a sequence that is oversampled by a factor of $L$ (due to the interpolation pre-processing) and thus bandlimited to only a bandwidth of $1/(LT_c)$. In light of this observation, the DPSS vectors can be used to reduce the dimension of the recovery problem.

A unique characteristic of the DPSS vectors is their ability to provide efficient and high quality approximations of oversampled discrete-time bandlimited signals. In particular, each row of $\boldvec{Z}$ (of length $r$) can be approximated with $k_D \approx 2N_DW_D$ expansion coefficients by computing inner products with a collection of $k_D$ DPSS vectors configured by setting $N_D=r$ and $W_D=1/(2L)$. In short, the use of such a DPSS vector-based dictionary provides a way for reducing the linear system of \eqref{eqn:finite_mc_dt_mat_vec_form} into a smaller set of equations (which is faster to solve), without sacrificing the rank of the measurement matrix. Define
\begin{equation*}
\boldvec{\widetilde{Z}}=\boldvec{ZQ},
%\label{eqn:dpss_measurements_coeff}
\end{equation*}
where $\boldvec{\widetilde{Z}}\in\mathds{C}^{q\times k_D}$ and $\boldvec{Q}\in\mathds{R}^{r\times k_D}$ is constructed as
\begin{equation*}
\boldvec{Q}\triangleq
\begin{bmatrix}
\boldvec{s}^{(0)}_{N_D,W_D}&\cdots&\boldvec{s}^{(k_D-1)}_{N_D,W_D}
\end{bmatrix},
%\label{eqn:dpss_dictionary}
\end{equation*}
with $\boldvec{s}^{(\ell)}_{N_D,W_D}$ as defined in \eqref{eqn:dpss_vector}. Accordingly, post-multiplying both sides of the linear system of \eqref{eqn:finite_mc_dt_mat_vec_form} gives
\begin{equation}
\boldvec{\widetilde{Z}}=\boldvec{A\widetilde{X}}_{\text{bb}},
\label{eqn:finite_mc_dt_mat_vec_form_dpss}
\end{equation}
where $\boldvec{\widetilde{X}}_{\text{bb}}=\boldvec{X}_{\text{bb}}\boldvec{Q}$. Thus, as an alternative to solving \eqref{eqn:finite_mc_dt_mat_vec_form}, one may solve the smaller system~\eqref{eqn:finite_mc_dt_mat_vec_form_dpss} instead and then finally compute
\begin{equation*}
\boldvec{X}_{\text{bb}}=\boldvec{\widetilde{X}}_{\text{bb}}\boldvec{Q}^T,
\end{equation*}
since the normalized columns of $\boldvec{Q}$ are orthogonal to each other.

We note that the linear of system of \eqref{eqn:finite_mc_dt_mat_vec_form_dpss} is an approximation of the linear system of \eqref{eqn:finite_mc_dt_mat_vec_form}. Therefore, depending on the application, if exact FH signal reconstruction is not necessary, one can potentially trade off FH signal reconstruction error for significant improvements in solver runtime latency by solving the linear system of \eqref{eqn:finite_mc_dt_mat_vec_form_dpss} instead of \eqref{eqn:finite_mc_dt_mat_vec_form}. In particular, the use of \eqref{eqn:finite_mc_dt_mat_vec_form} results in a reduction factor of $L$ (for $k_D=2N_DW_D$) in terms of the number of unknown $\boldvec{\widetilde{X}_{\text{bb}}}$ columns as compared to that of $\boldvec{X_{\text{bb}}}$.

Further, define the approximation error $\boldvec{E}_{\text{a}}\in\mathds{C}^{L\times r}$ associated with the approximation of the linear system of \eqref{eqn:finite_mc_dt_mat_vec_form} with that of \eqref{eqn:finite_mc_dt_mat_vec_form_dpss} as
\begin{equation}
\boldvec{E}_{\text{a}}\triangleq \boldvec{X}_{\text{bb}}-\boldvec{X}_{\text{bb}}QQ^T.
\label{def:mc_dpss_approx_error}
\end{equation}
The following lemma bounds this error.
\begin{lemma}
	The Frobenius norm of the approximation error $\boldvec{E}_{\text{a}}\in\mathds{C}^{L\times r}$ as defined in \eqref{def:mc_dpss_approx_error} satisfies
	\begin{equation*}
	||\boldvec{E}_{\text{a}}||_F\le \sqrt{L(\delta+rC_3e^{-C_4r})}||\boldvec{X}_{\text{bb}}||_F,
	\end{equation*}
	where constants $\delta$, $C_3$ and $C_4$ are as defined in Theorem~\ref{th:dpss_approx_quality_2}, provided the assumption in Theorem~\ref{th:dpss_approx_quality_2} is satisfied and $k_D=2N_DW_D(1+\epsilon)$.
	\label{lm:mc_dpss_approx_error}
\end{lemma}
\textbf{Proof}: See Appendix~\ref{adx:mc_dpss_approx_error}.

We note that the assumption in Theorem~\ref{th:dpss_approx_quality_2} is a mild condition on our $\afh{N}{B}{T}$ signal model that is mostly satisfied in practice. As discussed previously in Theorem~\ref{th:dpss_approx_quality_2}, in practice, one needs to use a value of $k_D$ that is slightly larger than $2N_DW_D$ for excellent approximation quality. We numerically demonstrate the dependence of the DPSS dictionary approximation quality on $k_D$ in Section~\ref{sec:mc_dpss_simulation}.

\section{Numerical Experiments}
\label{sec:mc_dpss_simulation}

\begin{figure}[t]
	\centering
	\begin{minipage}{0.48\textwidth}
		\centering
		\includegraphics[width=0.75\textwidth]{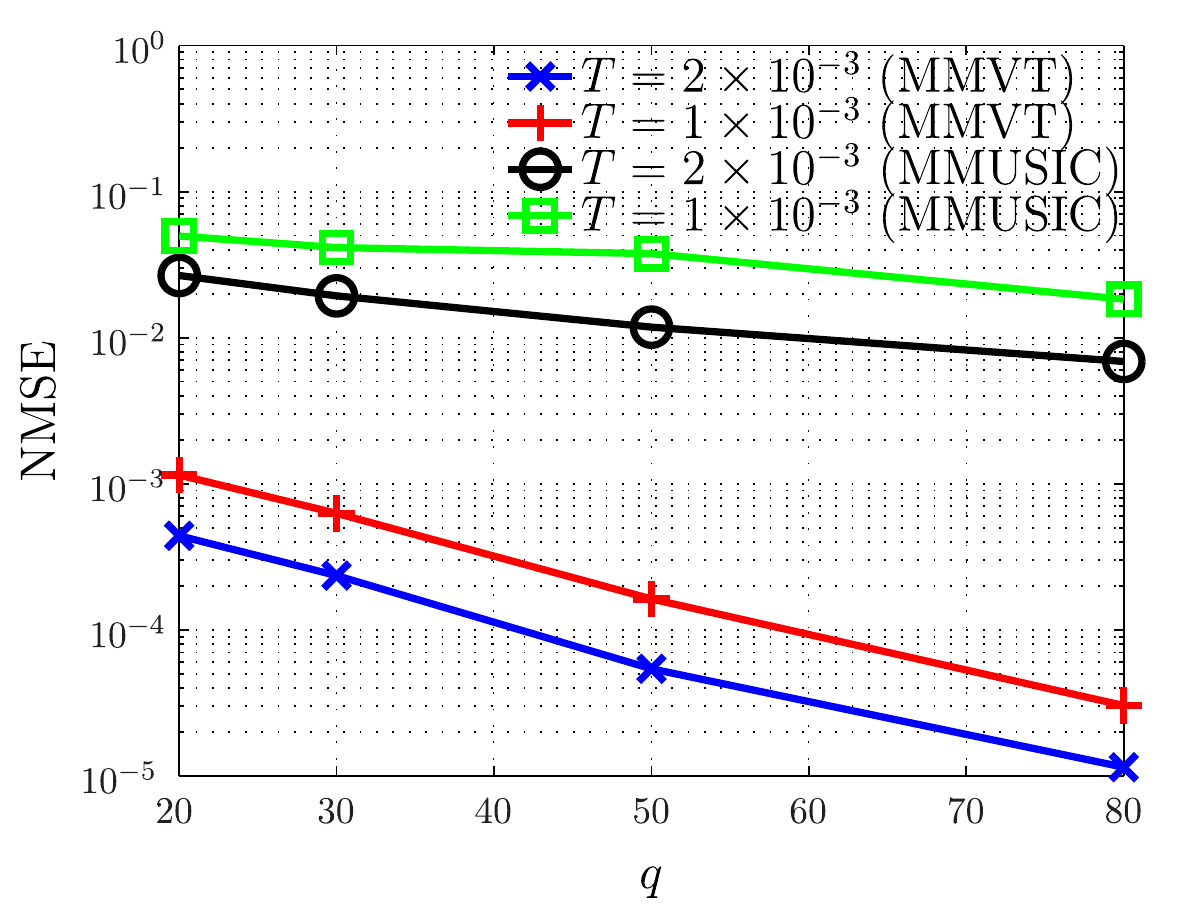}
		\label{fig:nmse_vs_q_500}
		\subcaption{{\small A plot of NMSE versus $q$ for an FH signal corresponding to $\afh{1}{25000}{T}$.}}
	\end{minipage}
	\hspace{1ex}
	\begin{minipage}{0.48\textwidth}
		\centering
		\includegraphics[width=0.75\textwidth]{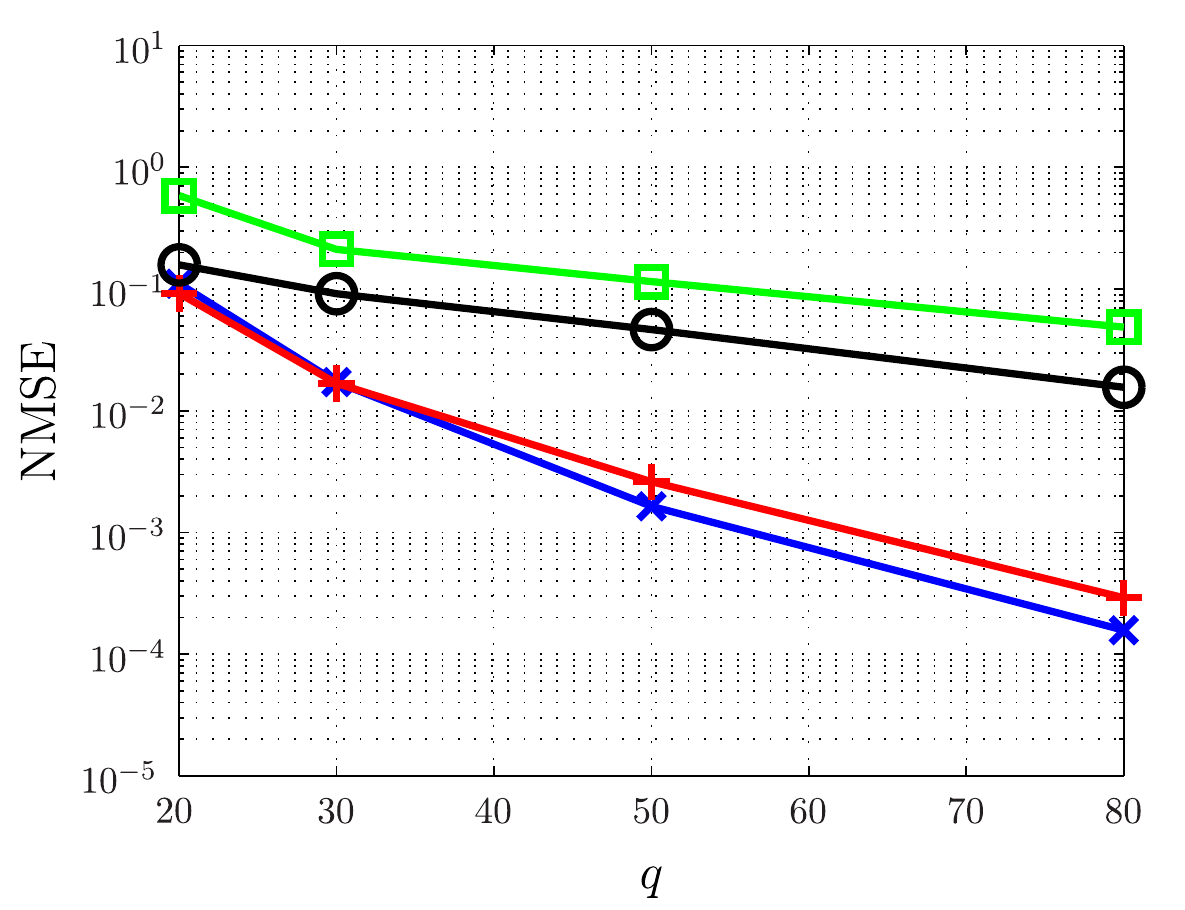}
		\label{fig:nmse_vs_q_1000}		
		\subcaption{{\small A plot of NMSE versus $q$ for a collection of FH signals corresponding to $\afh{7}{25000}{T}$.}}
		\end{minipage}
	\caption{\label{fig:nmse_vs_q}A comparison of NMSE versus $q$ for collections of FH signals corresponding to different $N$'s using the MMV-Time domain linear system of \eqref{eqn:finite_mc_dt_mat_vec_form} denoted by MMVT, and using the modified MUSIC algorithm denoted by MMUSIC. }
\end{figure}

We begin this section by discussing the $\afh{N}{B}{T}$ signal used in our numerical experiments. For all FH signals, the following time-limited and essentially bandlimited window was used:
\begin{equation*}
r(t) =
\begin{cases}
\frac{20}{T_H}t & \text{if } 0\le t< 0.05T_H, \\
1       & \text{if } 0.05\le t< 0.95T_H,\\
\frac{20}{T_H}(T_H-t)&\text{if } 0.95\le t<T_H,\\
0&\text{otherwise},
\end{cases}
\end{equation*}
where $T_H=0.95$HRI is the duration of one hop. In words, for each hop, $g_{ik}(t)$ ramps up $m_{ik}(t)$ for 5\% of the hop duration, maintains $m_{ik}(t)$ for 90\% of the hop duration and ramps down $m_{ik}(t)$ for 5\% of the hop duration. All FH hops were generated with a baseband waveform having 4PSK modulation type with pulses shaped using the raised cosine filter and configured with an excess bandwidth of 0.3. The symbol rate of $m_{ik}(t)$ was adjusted such that its resulting bandwidth was 25kHz. An $\afh{N}{25000}{T}$ signal was then generated by summing $N$ individual FH signals together. We note that all frequency hoppers were generated with identical HRIs equal to $T$, and this knowledge of $T$ was used to determine the appropriate duration of the segments used in the segment-based recovery framework. As we show (e.g., in Fig.~\ref{fig:nmse_vs_q}), one can expect improved signal recovery performance when the collection of FH signals contain other frequency hoppers operating at a larger HRI. An MC $\mc{4\times 10^{-7}}{100}{q}{C}$ sampler was used to sample the FH signal at sub-Nyquist rates which correspond to $q<L$ using a randomly chosen sampling pattern $C$. We note that $T_c=4\times 10^{-7}$ corresponds to a sampling rate of 2.5MHz for the final reconstructed $\afh{N}{25000}{T}$ signal.

We choose the SPGL1 MMV solver \cite{BergFriedlander:2008} as a representative CS MMV solver in our numerical experiments. The interested reader is referred to \cite{6621277} for a comparison of the performance of various recovery algorithms proposed previously for blind multiband signal reconstruction.

\begin{figure}[t]
	\centering
	\begin{minipage}{0.48\textwidth}
		\centering
		\includegraphics[width=0.75\textwidth]{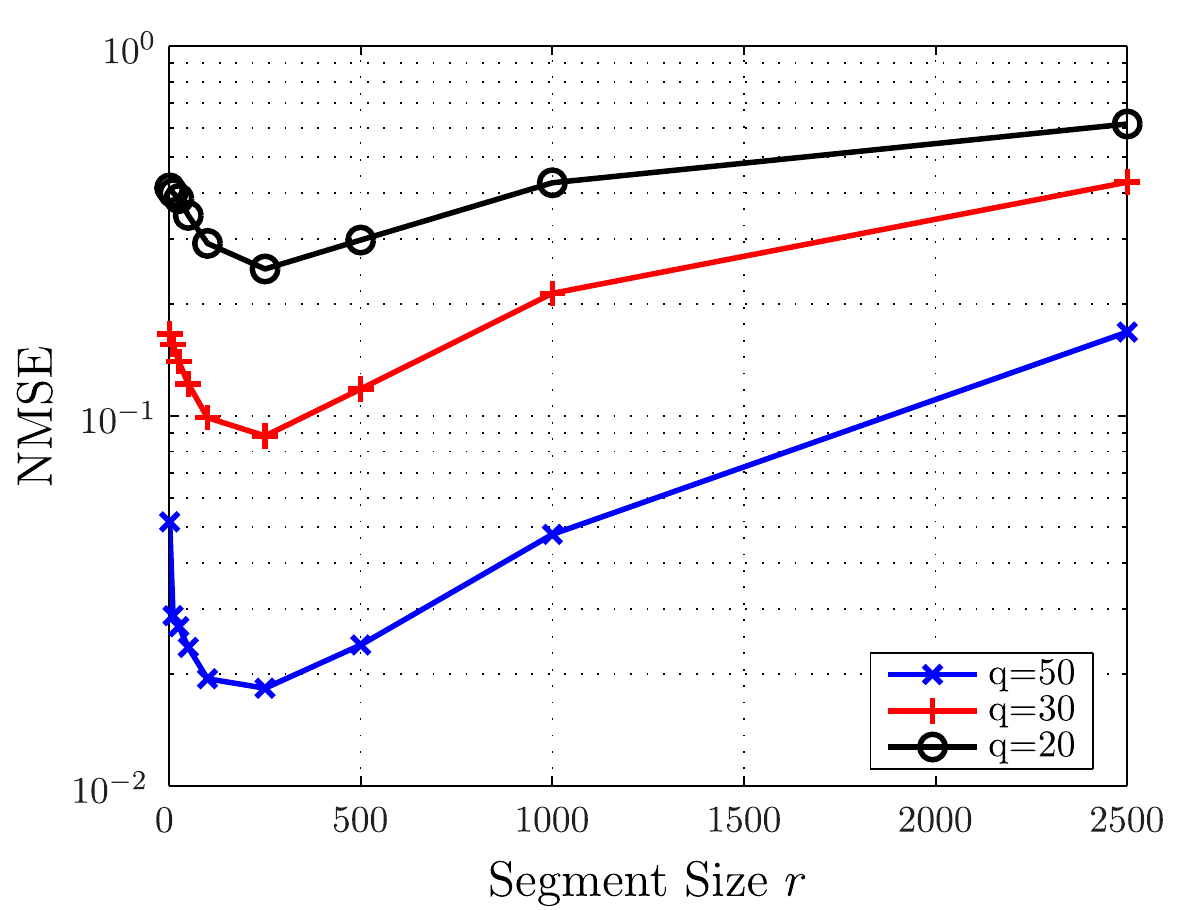}
		\subcaption{\label{fig:nmse_vs_r}A plot of NMSE versus segment size $r$ for a $\afh{7}{25000}{2\times 10^{-4}}$ signal for $q$=20, 30, 50, with corresponding effective compression levels $L/q$=5, 3.33 and 2.}
	\end{minipage}
	\hspace{1ex}
	\begin{minipage}{0.48\textwidth}
		\centering
		\includegraphics[width=0.75\textwidth]{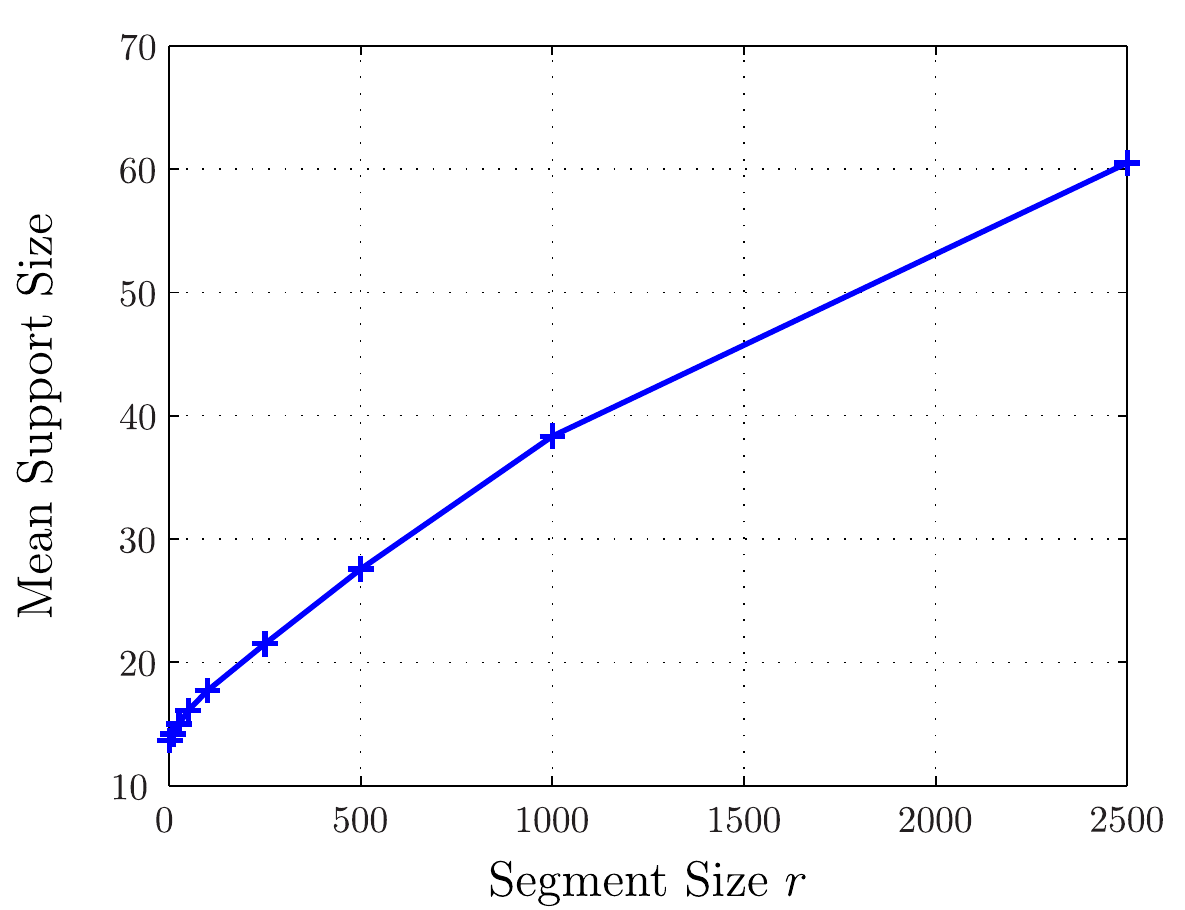}
		\subcaption{\label{fig:mean_supp_size_vs_r}A plot of mean support size versus segment size $r$ for the same $\afh{7}{25000}{2\times 10^{-4}}$ signal.}	\end{minipage}
	\caption{A plot of NMSE versus segment size $r$ for different $q$'s corresponding to a $\afh{7}{25000}{2\times 10^{-4}}$ and a plot of its corresponding mean support size versus segment size $r$.}
\end{figure}

In Section~\ref{sec:mc_dpss_blind_fh_reconstruction_mc_time_domain_equivalent_linear}, it was argued that choosing an appropriate value of the segment size $r$ for a given $\afh{N}{B}{T}$ signal amounts to a design trade-off. For various values of $r$, Fig.~\ref{fig:nmse_vs_r} shows the normalized mean square error (NMSE), defined as
\begin{equation*}
\text{NMSE}\triangleq\frac{||\hat{\boldvec{x}}-\boldvec{x}||^2_2}{||\boldvec{x}||^2_2},
%\label{def:nmse}
\end{equation*}
where $\hat{\boldvec{x}}$ denotes an estimate of the $\afh{N}{B}{T}$ signal and $\boldvec{x}$ denotes an observation interval of the $\afh{N}{B}{T}$ signal. We note that while a $\afh{N}{B}{T}$ signal with a duration of $100$ms was used, recovery of $\boldvec{X}_{\text{bb}}$ was performed over segments with varying $r$. Overall, an estimate of $\boldvec{X}_{\text{bb}}$ with duration 100ms was constructed by concatenating consecutive estimates of $\boldvec{X}_{\text{bb}}$ obtained from their corresponding segments.

Fig.~\ref{fig:nmse_vs_r} shows a plot of NMSE versus $r$ for various $q$ values (which scale proportionally to the average sampling rates of the MC sampler) using the MC segment-based recovery framework of \eqref{eqn:finite_mc_dt_mat_vec_form} and Fig.~\ref{fig:mean_supp_size_vs_r} shows its corresponding mean support size versus segment size $r$. The mean support size of the $\afh{7}{25000}{2\times 10^{-4}}$ signal is obtained by averaging the support size of the signal over all segments of size $r$ for a signal duration of 10ms. In particular, the underlying FH signal had an HRI of $0.2$ms (5000 hops/s), which is equivalent to a segment size of 500 (for $T_c=4\times10^{-7}$s). We note that the plots in Fig.~\ref{fig:nmse_vs_r} indicate that in the regime when $r\ll \text{round}(T/T_c)$ (i.e., the segment size is small relative to the HRI of the FH signal), increasing $r$ decreases the overall NMSE of the recovered signal. This is due to the effect of increasing $\text{rank}(\boldvec{Z})$ which improves the MMV solver performance due to the increased ability of finding the true support of $\boldvec{X}_{\text{bb}}$. On the other hand, increasing $r$ too much results in an increasing NMSE due to the increasing number of nonzero rows of $\boldvec{X}_{\text{bb}}$ as discussed previously. Due to these opposing effects, the plots in Fig.~\ref{fig:nmse_vs_r} suggest an optimum $r\approx \text{round}(T/(2T_c))$, which can be explained intuitively. In particular, one would expect an optimum value of $r\approx \text{round}(T/T_c)$ as this is the smallest segment size for which the fastest frequency hoppers do not switch frequency. However, due to the asynchronous nature of the frequency hoppers, most frequency hoppers will change frequency once during an arbitrary segment of size $r\approx \text{round}(T/T_c)$. Setting $r\approx \text{round}(T/(2T_c))$ guarantees that, on average, some of the hoppers do not change frequency during any given segment. This is corroborated with the plot in Fig.~\ref{fig:mean_supp_size_vs_r} showing the moderate increase (much less than doubling) in the mean size of the signal's support as the segment size increases from 250 to 500.

We next examine the impact of the parameters $N$, $T$, as well as the number of channels ($q$=20, 30, 50 and 80) on the overall $\afh{N}{B}{T}$ signal reconstruction quality in terms of NMSE. In these and subsequent numerical experiments, we have used $r=\text{round}(T/(2T_c))$.

Fig.~\ref{fig:nmse_vs_q} shows a comparison of the NMSE versus $q$ for collections of FH signals corresponding to $N=1,7$ with HRIs corresponding to 500 hops/s and 1000 hops/s using the MMV-Time domain linear system of \eqref{eqn:finite_mc_dt_mat_vec_form} denoted by MMVT (blue and red curves) and the modified MUSIC algorithm denoted by MMUSIC (black and green curves). We note that in general, an increase in $q$ leads to a decrease in NMSE and hence, an improvement in signal reconstruction quality.  As the HRIs decrease (i.e., increasing hopping rates), the increase in NMSE (for a fixed $q$) can be attributed to the increase in number of hops thereby increasing the number of nonzero rows of $\boldvec{X}_{\text{bb}}$. We note that with each switching of frequency, additional rows in $\boldvec{X}_{\text{bb}}$ become nonzero due to the transient signal amplitude changes involved as the FH signal ramps up or down. When $N=7$, the increase in NMSE (due to switching) is less obvious since NMSE is largely dominated by the effect of a large $N$. In addition, NMSE versus $q$ plots (the black and green curves) corresponding to FH signal reconstruction using the modified MUSIC algorithm, denoted by MMUSIC, show a consistent poorer FH signal reconstruction quality. We note that in this setup, the MMUSIC algorithm was used to estimate the support of each segment of the collection of FH signals. Let us remind the reader that for the $\afh{N}{B}{T}$ signal model, support recovery using the MMUSIC algorithm is not expected to work well as it consistently underestimates the size of the support of the collection of FH signals. As such, subsequent FH signal reconstruction on an underestimated support size accounts for the significantly higher NMSE. 

\begin{figure}[t]
	\centering
	\begin{minipage}{0.48\textwidth}
		\centering
		\includegraphics[width=0.75\textwidth]{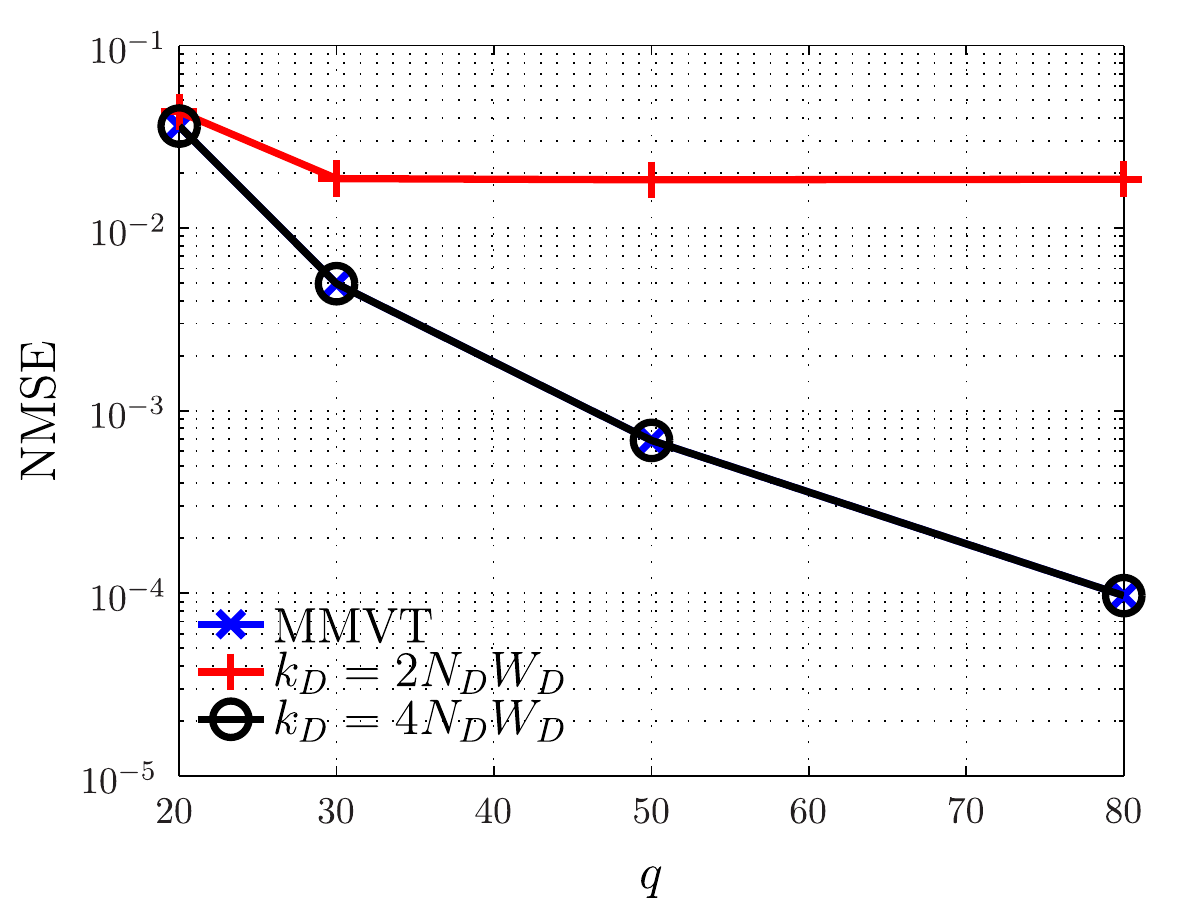}
		\label{fig:nmse_vs_q_dpss_500}
		\subcaption{{\small A plot of NMSE versus $q$ for a collection of FH signals corresponding to $\afh{5}{25000}{2\times 10^{-3}}$.}}
	\end{minipage}
	\hspace{1ex}
	\begin{minipage}{0.48\textwidth}
		\centering
		\includegraphics[width=0.75\textwidth]{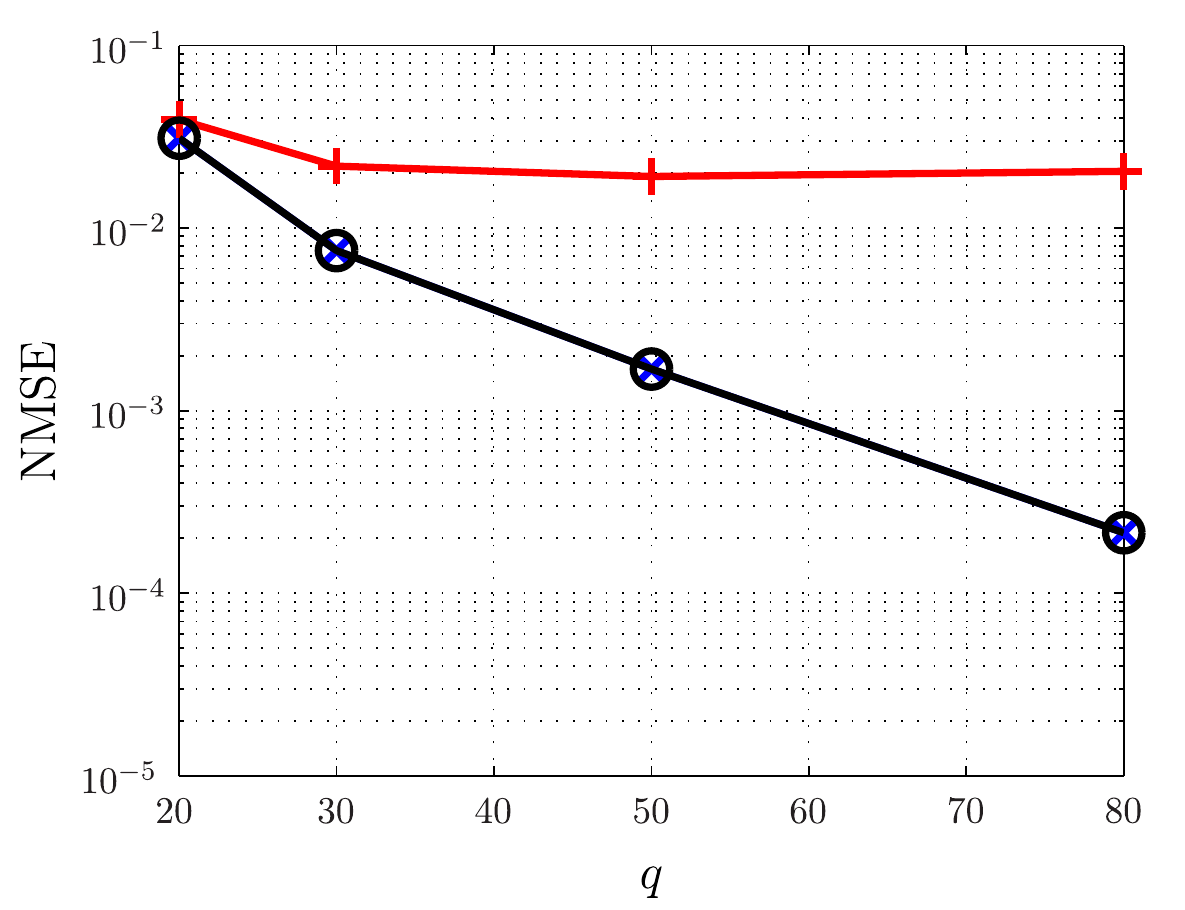}
		\label{fig:nmse_vs_q_dpss_1000}
		\subcaption{{\small A plot of NMSE versus $q$ for a collection of FH signals corresponding to $\afh{5}{25000}{1\times 10^{-3}}$.}}
	\end{minipage}
	\caption{\label{fig:nmse_vs_q_dpss}A comparison of NMSE versus $q$ for collections of signals corresponding to different $T$'s using the MMV-Time domain linear system of \eqref{eqn:finite_mc_dt_mat_vec_form} denoted by MMVT, and that of \eqref{eqn:finite_mc_dt_mat_vec_form_dpss} using the DPSS dictionary with $k_D=2N_DW_D$ and $4N_DW_D$.}
\end{figure}

Fig.~\ref{fig:nmse_vs_q_dpss} shows a comparison of NMSE versus $q$ for collections of signals recovered using the MC segment-based recovery framework of \eqref{eqn:finite_mc_dt_mat_vec_form} and that of \eqref{eqn:finite_mc_dt_mat_vec_form_dpss} using the DPSS dictionary previously discussed in Section~\ref{sec:mc_segment_recovery_dpss}. We note that the red curves ($k_d=2N_DW_D$) in both plots of Fig.~\ref{fig:nmse_vs_q_dpss} show a nondecreasing trend even with an increasing $q$, but further increasing $k_D$ to $4N_DW_D$ (black curves) gives a resulting NMSE that closely matches that when solving the linear system of \eqref{eqn:finite_mc_dt_mat_vec_form}. As discussed previously, in practice one needs to use a value of $k_D$ slightly larger than $k_d=2N_DW_D$ to see excellent approximation quality.

\begin{figure}[t]
	\centering
	\begin{minipage}{0.48\textwidth}
		\centering
		\includegraphics[width=0.75\textwidth]{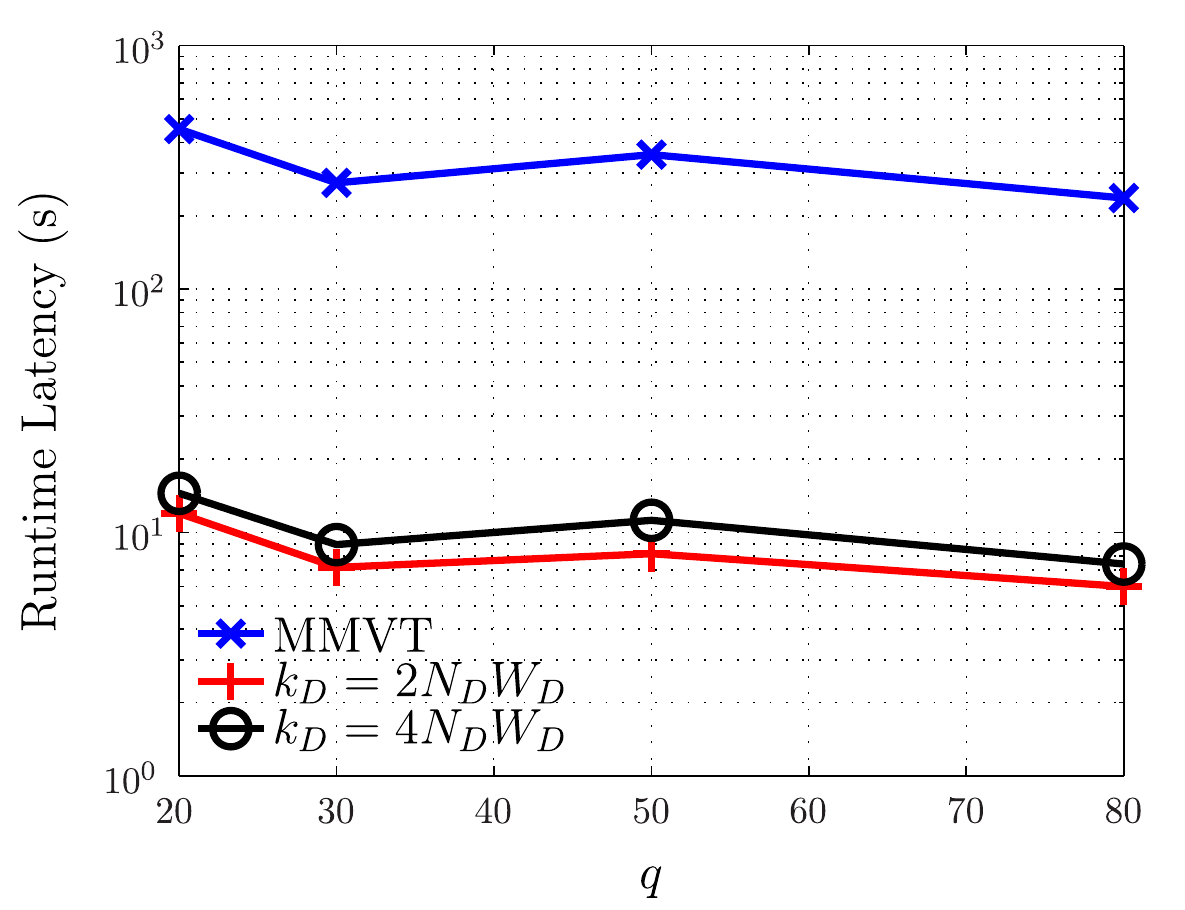}
		\label{fig:rl_vs_q_dpss_500}
		\subcaption{{\small A plot of runtime latency versus $q$ for a collection of FH signals corresponding to $\afh{5}{25000}{2\times 10^{-3}}$.}}
	\end{minipage}
	\hspace{1ex}
	\begin{minipage}{0.48\textwidth}
		\centering
		\includegraphics[width=0.75\textwidth]{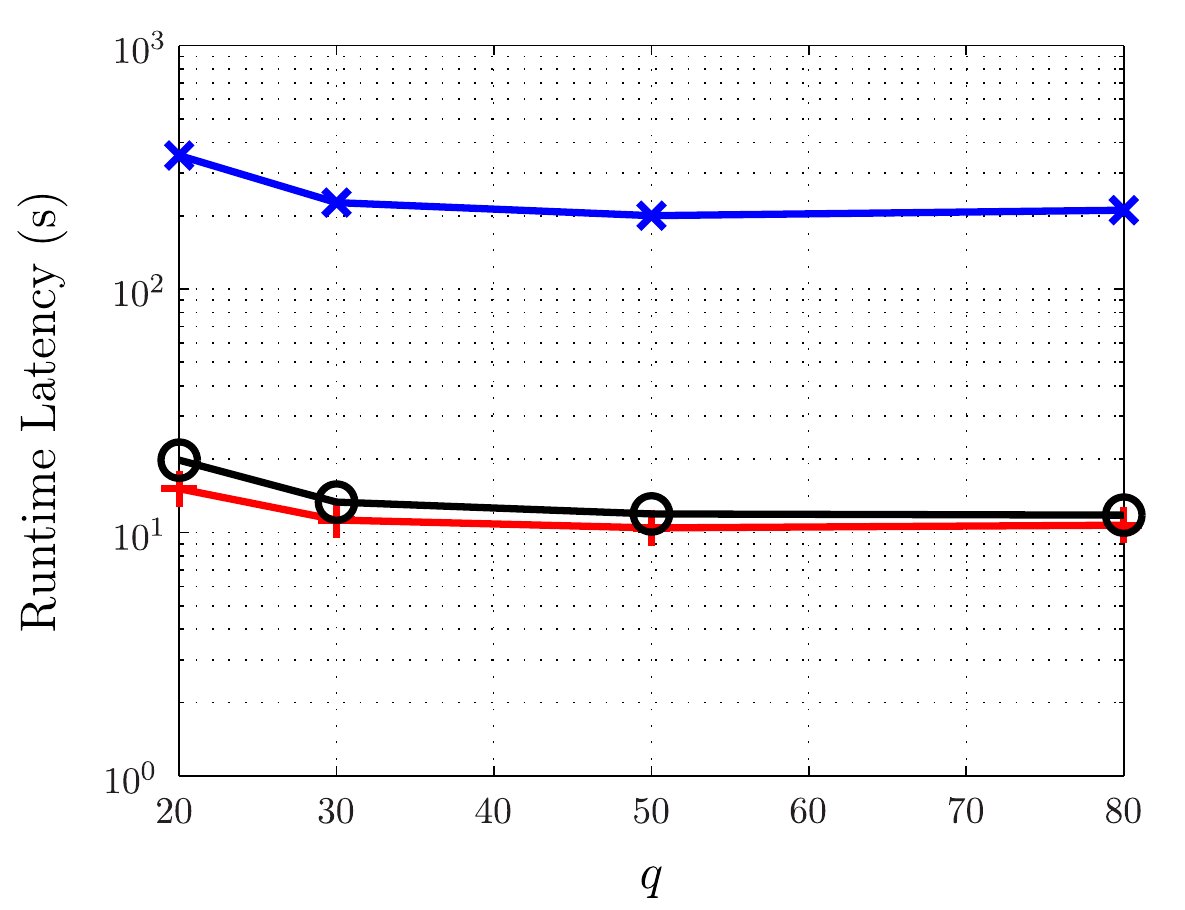}
		\label{fig:rl_vs_q_dpss_1000}
		\subcaption{{\small A plot of runtime latency versus $q$ for a collection of FH signals corresponding to $\afh{5}{25000}{1\times 10^{-3}}$.}}
	\end{minipage}
	\caption{\label{fig:rl_vs_q_dpss}A comparison of runtime latency (s) versus $q$ for collections of signals corresponding to different $T$'s using the MMV-Time domain linear system of \eqref{eqn:finite_mc_dt_mat_vec_form} denoted by MMVT, and that of \eqref{eqn:finite_mc_dt_mat_vec_form_dpss} using the DPSS dictionary with $k_D=2N_DW_D$ and $4N_DW_D$.}	
\end{figure}

In Fig.~\ref{fig:rl_vs_q_dpss}, we provide an empirical runtime latency comparison when using the DPSS dictionary (red and black curves) as compared to solving the linear system of \eqref{eqn:finite_mc_dt_mat_vec_form} (blue curves). The curves in Fig.~\ref{fig:rl_vs_q_dpss} indicate a $10\times$ reduction in in runtime latency for our numerical experiment setup. Let us remind the reader that the use of the DPSS dictionary allows one to reduce the scale of the recovery problem due to the reduction of the number of columns of unknown vectors from $r$ to $k_D$. We note that this reduction is by a factor of $1/L$ and $2/L$ when $k_D=2N_DW_D$ and $4N_DW_D$, respectively.

\section{Conclusion}
\label{sec:mc_dpss_conclusion}

For blind recovery of FH signals from MC samples, it is necessary to consider a segment-based recovery framework. In this paper, we have highlighted a salient aspect of the MC sampling protocol, namely that the MC discrete-time equivalent linear measurement system facilitates segmentation into finite-dimension problems that fit neatly into the conventional CS MMV framework. Of particular importance is the fact that the time-varying spectrum of the FH signals will yield sparse supports in the MMV problem as long as the segments are chosen to be sufficiently short in duration. Large-scale simulations confirm the effectiveness and viability of this framework. We have also proposed the use of a dictionary comprised of DPSS vectors for reducing the dimension of the recovery problem, and we have demonstrated that this can lead to significant computational savings. Our work is an extension of previous works focusing on the multiband signal model and could potentially enable sub-Nyquist rate sampling of other signals, which are sparse in time-frequency plane, using a MC sampler.

There are open questions not yet addressed in this paper. Due to the multi-channel architecture of the MC sampler, we did not address issues associated with non-identical channel characteristics such as non-ideal ADCs and imperfect clock synchronization resulting in incorrect sampling instants across channels. However, we believe there exist calibration techniques in the field of time-interleaved ADCs which could possibly address this issue.

In our numerical experiments, we did not investigate the impact of nonuniform FH baseband signal bandwidths on the sampling rate reduction; rather we have used a fixed baseband signal bandwidth of 25kHz. We do believe there exist algorithms from previous works \cite{4601017,4749297,544131,868487,950786} focusing on multiband signals which can be incorporated into the MC segment-based recovery framework to address this issue.

\section*{Acknowledgments}
\label{sec:ack}

We are grateful to the authors of \cite{lexa_1} and \cite{BergFriedlander:2008} for making available the Continuous-Time Spectrally-Sparse (CTSS) Sampling Toolbox and the SPGL1 MMV Solver, respectively, which helped to shorten the development time of the codes used in our numerical experiments.

\appendix
\subsection{Proof of Lemma~\ref{lm:mc_dt_mat_vec_form} }
\label{adx:mc_dt_mat_vec_form}

From the definition of $z_i(k)$ in \eqref{eqn:interpolate_offset_correct_mc_output} it follows that the DTFT of $z_i(k)$ will equal
\begin{equation*}
	\dtft{Z_i}{T_c} =
	\begin{cases}
	e^{-j2\pi fc_iT_c}\dtft{Y_i}{LT_c}, & %f\in\left[\frac{1}{LT_c}\left(n-\frac{1}{2}\right),\frac{1}{LT_c}\left(n+\frac{1}{2}\right)\right],\: n\in\mathds{Z}, \\
	f\in\mathcal{F}_0,\\
	0,       & \text{otherwise.}
	\end{cases}
\end{equation*}
Let $\dtft{X_{\ell,\text{bb}}}{T_c}$ be as defined in~\eqref{eqn:timedomainequivDTFTxbb}, and let
\begin{equation*}
	\boldvec{z}(f)\triangleq
	\begin{bmatrix}
	\dtft{Z_1}{T_c}&\cdots&\dtft{Z_q}{T_c}
	\end{bmatrix}^T\in\mathds{C}^q
\end{equation*}
and
\begin{equation*}
	\boldvec{x}_{\text{bb}}(f)\triangleq
	\begin{bmatrix}
	\dtft{X_{0,\text{bb}}}{T_c}&\cdots&\dtft{X_{L-1,\text{bb}}}{T_c}
	\end{bmatrix}^T\in\mathds{C}^L,
\end{equation*}
with corresponding inverse DTFT $\boldvec{x}_\text{bb}(k) \in \mathds{C}^L$. Due to \eqref{eqn:mc_mat_vec_form},
\begin{equation}
	\boldvec{z}(f)=
	\begin{cases}
	\boldvec{Ax}_{\text{bb}}(f),& %f\in\left[\frac{1}{LT_c}\left(n-\frac{1}{2}\right),\frac{1}{LT_c}\left(n+\frac{1}{2}\right)\right],\: n\in\mathds{Z},\\
	f\in\mathcal{F}_0,\\
	0,&\text{otherwise}.
	\end{cases}
	\label{pf:dtft_to_dtft_T_c}
	\end{equation}
since the entries of $\boldvec{z}(f)$ and $\boldvec{x}_{\text{bb}}(f)$ are identical to the entries of $\boldvec{y}(f)$ and $\boldvec{x}(f)$, respectively for $f\in \mathcal{F}_0$. Applying the inverse DTFT to both sides of \eqref{pf:dtft_to_dtft_T_c} completes the proof.

\subsection{Proof of Corollary~\ref{lm:finite_mc_dt_unique_sol}}
\label{adx:finite_mc_dt_unique_sol}

As previously discussed in Section~\ref{sec:mc_dpss_cs_prelim}, a known condition which guarantees the existence of a unique solution to \eqref{eqn:finite_mc_dt_mat_vec_form} (assuming an MMV problem) is
\begin{equation*}
||\boldvec{X}_{\text{bb}}||_0<\frac{\text{spark}(\boldvec{A})+\text{rank}(\boldvec{Z})-1}{2},
%\label{pf:finite_mc_dt_unique_sol_1}
\end{equation*}
where $||\boldvec{X}_{\text{bb}}||_0$ denotes the number of nonzero rows of $\boldvec{X}_{\text{bb}}$. By make the following substitutions, $||\boldvec{X}_{\text{bb}}||_0=4N$, and $\text{spark}(\boldvec{A})=q+1$, followed by trivial algebraic manipulations, one completes the proof.

\subsection{Proof of Lemma~\ref{lm:mc_dpss_approx_error}}
\label{adx:mc_dpss_approx_error}

Assuming the conditions of Theorem~\ref{th:dpss_approx_quality_2} are satisfied, the approximation error (when using $\boldvec{Q}$) for each row of $\boldvec{X}_{\text{bb}}$, denoted by $\boldvec{x}_i$ can be bounded by invoking Theorem~\ref{th:dpss_approx_quality_2}:
\begin{equation}
	||\boldvec{x}_i-\boldvec{P_Q}\boldvec{x}_i||^2_2\le (\delta+rC_3e^{-C_4r})||\boldvec{x}_i||^2_2,\quad\forall\;i=1,\dots,L,
	\label{pf:dpss_row_bound}
\end{equation}
where $\boldvec{P_Q}=\boldvec{QQ}^T$ and constants $\delta$, $C_3$ and $C_4$ are as defined in Theorem~\ref{th:dpss_approx_quality_2}. By summing \eqref{pf:dpss_row_bound} for all $L$ rows, one arrives at
\begin{equation}
	||\boldvec{X}_{\text{bb}}-\boldvec{X}_{\text{bb}}\boldvec{QQ}^T||_F^2\le L(\delta+rC_3e^{-C_4r})||\boldvec{X}_{\text{bb}}||^2_F,
	\label{pf:dpss_approx_error_square}
\end{equation}
and one completes the proof by taking square root on both sides of \eqref{pf:dpss_approx_error_square}.

\bibliographystyle{unsrt}
\bibliography{mc_dpss_ref}

\end{document}